\documentclass[12pt]{article}
\usepackage[natbibapa]{apacite}
\usepackage{amsmath}
\usepackage[dvipsnames]{xcolor}
\usepackage[
colorlinks=true,
citecolor=gray
]{hyperref}

\usepackage[margin=1in]{geometry}
\usepackage{times}
\usepackage{graphicx}
\usepackage{lineno}
\usepackage{parskip}
\usepackage{multicol}
\usepackage{enumitem}
\usepackage[doublespacing]{setspace}

\usepackage{wordcount}

\usepackage{listings}
\lstdefinelanguage{webppl}{
  keywords={var,sample,factor,condition,return,Infer,function,uniformDraw},
  morecomment=[l]{//},
}
\usepackage{simplebnf}
\usepackage{mathpartir}

\title{Theories of Mind as \\ Domain-Specific Languages of Thought}
\author{
\textbf{Kartik Chandra} \\
Massachusetts Institute of Technology \\
Department of Electrical Engineering and Computer Science (EECS) \\
Computer Science and Artificial Intelligence Laboratory (CSAIL) \\
Cambridge, MA, USA \\
\texttt{kach@mit.edu}
\and
\textbf{Jonathan Ragan-Kelley} \\
Massachusetts Institute of Technology \\
Department of Electrical Engineering and Computer Science (EECS) \\
Computer Science and Artificial Intelligence Laboratory (CSAIL) \\
Cambridge, MA, USA \\
\texttt{jrk@mit.edu}
\and
\textbf{Joshua B.\ Tenenbaum} \\
Massachusetts Institute of Technology \\
Department of Brain \& Cognitive Sciences (BCS) \\
Computer Science and Artificial Intelligence Laboratory (CSAIL) \\
Cambridge, MA, USA \\
\texttt{jbt@mit.edu}
\and
\textbf{Rebecca Saxe} \\
Massachusetts Institute of Technology \\
Department of Brain \& Cognitive Sciences (BCS) \\
McGovern Institute for Brain Research (MIBR) \\
Cambridge, MA, USA \\
\texttt{saxe@mit.edu}
}

\date{September 2026}

\begin{document}

\singlespacing
\maketitle

\newpage

\paragraph{Short abstract:}
\begin{wordcount}{\shortabstractwords}
What kind of thing is a ``theory of mind''? We propose to formalize theories of mind as domain-specialized programming languages, which can be used to reason about the mental states of other agents. This perspective allows us to apply ideas from programming language theory to lend insight into a variety of theoretical issues in theory-of-mind research: how a single, fixed intuitive theory can be used to reason about social situations with unbounded structural variation; how a modular, domain-specialized cognitive system can interface with external world knowledge; and how our theories of mind might grow over the course of development.
\end{wordcount}

\paragraph{Long abstract:}
\begin{wordcount}{\longabstractwords}
What kind of thing is a ``theory of mind''? We propose to formalize theories of mind as domain-specialized programming languages, which can be used to reason about the mental states of other agents. Our proposal builds on the longstanding idea that thinking is akin to programming in an internal ``language of thought.'' But rather than positing a single general-purpose language for all of thought, we posit a collection of domain-specific languages of thought, whose respective syntaxes and semantices encode the theories of various domains.
We make this idea concrete in the domain of intuitive psychology using a real-world programming language called memo, which is specialized for theory-of-mind reasoning via domain-specialized syntactic constructs like ``knows,'' ``wants,'' and ``thinks.'' In our view, using one's theory-of-mind to think about a social situation is analogous to using memo to write a program that models that situation.
Through a series of worked examples, we show how formally representing a theory of mind as a memo-like programming language can lend insight into a variety of theoretical issues in theory-of-mind research: (1)~how a single, fixed intuitive theory can be used to reason about social situations with unbounded structural variation; (2)~how a modular, domain-specialized, and informationally-encapsulated system for social cognition might interface with external world knowledge, as well as other cognitive systems like intuitive physics; and (3)~how different theories of mind, such as those of children and adults, differ, what exactly is gained as theories grow over the course of development, and how that growth might occur.
\end{wordcount}


% $ texcount -inc -incbib -merge tomalotmore.tex ./.texpadtmp/tomalotmore.bbl
\paragraph{Word counts:}
Short abstract: \shortabstractwords;
Long abstract: \longabstractwords;
Main text: 14,877;
References: 7,309;
Total: 24,041

\newpage
\tableofcontents

\doublespacing

\section{What kind of thing is a theory of mind?}

When \citet{premack1978does} asked, ``Does the chimpanzee have a theory of mind?'' their use of the indefinite article \emph{``a''} launched an enduring controversy about how we should conceive of the capacity for reasoning about other minds. Their wording raised the possibility that even quite different ways of reasoning about other minds could be members of the same category: a ``theory of mind'' is a \emph{kind of thing}, of which the chimpanzee may possess an instance, which may differ from the instances possessed by a typical human adult or a child.

If this is the case, then what kind of thing might a theory of mind be? On the influential ``theory-theory'' \citep{gopnik1997words, carey1985conceptual}, a theory of mind is, indeed, a theory, akin to scientific theories \citep{gopnik1992child}. But this proposal has been contentious. A key problem is that it is not obvious that there could be any ``theory'' that meets all of the desiderata demanded of it by the theory-theorist. How could a single, finite body of knowledge be used to reason flexibly about all of the different social situations that people can reason about? How could an abstract, bounded, domain-specialized theory be applied to reason about real-world social situations, which require unfettered access to all world knowledge \citep{fodor1983modularity, garfield2001social, currie2000think}? What type of learning mechanism could produce an adult-like theory of mind starting from a child-like theory of mind, and how could such a learning mechanism explain the child's typical developmental trajectory \citep{goldman2006simulating}?

Debates around these questions have persisted for decades, in part because hypotheses about the nature of intuitive theories are typically articulated only qualitatively, in natural language. Even with the most careful exposition, a theory of theories expressed in words leaves room for imprecision and reinterpretation: we can only get so far by saying that a theory of mind is a kind of ``theory.'' What kind of thing is a theory, then? What knowledge is and isn't built into the theory? How can we tell if a line of reasoning is consistent with a theory? How can we tell if a given problem is within the domain of a theory at all? How can we tell whether two theories are the same or different? How can we tell if one theory is incommensurate with another? Recognizing the difficulty in unambiguously answering these questions in words, cognitive scientists have long endeavored to formalize theories of mind in precise mathematical terms instead.

The first step in this project is to choose a mathematical framework with which to formalize theories of mind. Suppose we claim that some entity (a human adult, a child, a chimpanzee, a large language model) reasons about other minds by applying theory-of-mind $\theta$. What kind of mathematical object is $\theta$? Should we think of $\theta$ as a set? A matrix? A function? A collection of propositions? The choice of mathematical object is not a mere implementation detail or technicality. Rather, this choice determines how we think about what kind of thing a theory of mind is: how it is represented, what operations and affordances it supports, and how it may be learned.

To date, the dominant approach to formalizing theories of mind has been to represent them as mathematical objects known as \emph{probabilistic models} \citep{gopnik2004theory}. This approach is appealing because it allows us to codify the causal knowledge of an intuitive theory using the machinery of causal Bayesian networks \citep{pearl2000causality, glymour2001mind} and their generalizations, which support key theory-like operations such as prediction, planning, inference, and counterfactual reasoning. For example, we might formalize an intuitive belief-desire psychology as a Bayesian network with nodes for belief and desire that are causally linked to a node for action; this Bayesian network could then be used to predict action given belief and desire, to infer belief and desire given action, and so on. Indeed, computational cognitive scientists have used this approach to concretely implement probabilistic models that make precise, quantitative predictions about people's intuitions in a wide variety of theory-of-mind tasks, ranging from inferring agents' goals and beliefs from their actions \citep{baker2009action, baker2017rational}, to moral reasoning \citep{kleiman2017learning}, emotion understanding \citep{houlihan2023emotion} pragmatic language use \citep{frank2012predicting}, social planning \citep{ho2022planning}, and more \citep[see][for a review]{cushman2024computational}. In this way, thinking of theories of mind as probabilistic models has been tremendously influential, and this view has enabled great progress in reverse-engineering the computations performed by theory-of-mind reasoning.

Yet, as \citet{tenenbaum2007intuitive} observe, this view of theory-of-mind is incomplete in an important way. Every probabilistic model $\mu_i$ that has been built to date formalizes a hypothesis about how people apply their theory of mind to reason about a particular agent engaged in a particular task: a graduate student foraging among food trucks, a strategic player playing tic-tac-toe. No probabilistic model is a model of a theory-of-mind \emph{itself}, in the way that no individual circuit diagram captures the theory of electromagnetism itself and no individual free-body diagram captures the theory of Newtonian mechanics itself. A theory $\theta$ is a more abstract and flexible object than a particular model $\mu_i$ that is instantiated by applying that theory to a given situation. To address the debates we raised at the beginning of this paper, we need a mathematical framework for formalizing the abstract theory $\theta$, an account of how the mind applies theory $\theta$ (in conjunction with the relevant external information) to generate concrete situation-specific models $\{\mu_1, \mu_2, \mu_3, \dots\}$, and an account of how $\theta$ itself can change over time. As we discuss in Section~\ref{sec:structural-variation}, it is not clear that probabilistic models on their own offer the best mathematical language to give that account in. Certainly, probabilistic models on their own have to date been unable to settle debates on the structure, modularity, and development of theory of mind.

The goal of this paper is to propose using a very different type of mathematical language to think about theories of mind: a mathematical language that provides the vocabulary we need to begin to formally address these more abstract issues related to structure, modularity, and development. Our view is consistent with but complementary to the probabilistic modeling tradition. We seek a type of mathematical object that is compatible with thinking of situation-specific reasoning in terms of probabilistic models, but can in addition meet at least three new challenges:

\subsection{The challenges of formalizing theories-of-mind}\label{sec:challenges}

The first challenge is to formalize in a single, fixed mathematical object the shared theory-of-mind $\theta$ behind all situation-specific theory-of-mind models $\mu_i$. What makes this challenging is that the mathematical object $\theta$ must somehow support unbounded structural variation in the types of situations that people can apply $\theta$ to reason about: this includes variation in the number of agents, the number of levels of recursive mentalizing those agents engage in, and the types of mental states they represent. A single Bayesian network, for example, cannot capture this type of structural variation, because it has a fixed number of nodes and edges.

The second challenge is for $\theta$ to only formalize theory-of-\emph{mind}, no more, no less. For example, any existing theory-of-mind model that reasons about an agent navigating a maze encodes within it some knowledge of impenetrability of walls, and any existing computational model that reasons about an agent playing tic-tac-toe encodes within it knowledge of the rules of tic-tac-toe. But none of this knowledge should be considered part of our theory of ``mind'' (though some of it might be part of other intuitive theories; for example, impenetrability may be a part of intuitive physics). The challenge is for $\theta$ to formalize only the parts of the model that are theory-of-mind, in a way that can nonetheless interface with extratheoretic world knowledge in order to build models of any task that humans can reason about. As we will discuss below, a solution that meets this challenge would carry implications for how we think about the modularity of social cognition in the mind.

The third challenge is for this mathematical framework to be flexible enough to formalize not only $\theta$, a theory-of-mind representative of a typical adult human, but also alternate theories of mind. For example, we would like to be able to use the same mathematical framework to formalize $\theta_3$, the theory-of-mind that we might hypothesize is held by a three-year-old child who cannot represent false beliefs. Formalizing alternate theories of mind in a common mathematical language would be valuable because it would allow us to precisely state and compare our hypotheses about theories-of-mind: for example, to compare the theories of mind hypothesized to be held by adults, by children at different developmental stages, by non-human primates, and by AI systems. Particularly in the case of development, this would also allow us to formalize hypotheses about how theories-of-mind are learned. For example, we would like to be able to compare $\theta_3$ and $\theta$ in order to precisely state what a child gains when she gains the ability to represent false beliefs, and we would like to be able to formally articulate hypotheses about how that child might make the developmental transition from $\theta_3$ to $\theta$.

Note that we seek to address these challenges by modeling theories of mind at the abstract level of cognitive representations. A parallel approach is to model theory-of-mind reasoning at the neural level, for example by training artificial neural networks on theory-of-mind tasks \citep{rabinowitz2018machine, gandhi2021baby} or by searching for theory-of-mind-specific circuits in pre-trained foundation models \citep{alkhamissi2025llm, tsvilodub2026emergent}. The neural network approach is different from but complementary to our approach: given a cognitive representation, we can ask how it could be implemented in a neural network model, and conversely, given a neural network model, we can ask which cognitive representations it implements \citep{piantadosi2024concepts}.

\subsection{Our proposal: theories of mind as programming languages}

What mathematical framework could address these challenges? We propose to formalize a theory of mind, not as a single probabilistic model, but rather as a \emph{programming language} that can be used to program any task-specific probabilistic model. This programming language is not a general-purpose language like Python or C++, but rather a ``domain-specific'' programming language (a DSL; Box~1) which is specialized for programming probabilistic models that reason about other minds. Rather than being organized around general-purpose syntactic forms like ``if'' and ``while,'' a DSL for reasoning about other minds might be organized around specialized syntactic forms like ``knows'' and ``wants,'' which can be composed generatively to express situations with arbitrarily many agents and recursive levels of reasoning (``Alice knows that Bob wants...''). These specialized syntactic forms, and their associated semantics, define the space of possible theory-of-mind models that can be programmed using this DSL.

On our view, the relationship between a DSL and the programs that can be built by using that DSL is analogous to the relationship between a theory and the models that can be built by applying that theory. Models and programs represent reasoning about particular situations, while theories and DSLs encode abstract principles that transcend any particular situation. DSLs however additionally come equipped with a well-established means of being formalized precisely as mathematical objects: a DSL is defined by its syntax and semantics, which in turn can be formally specified using tools from the discipline of programming language theory \citep{pierce2002types, felleisen2009semantics}. By associating intuitive theories with DSLs, we can apply these tools to formalize intuitive theories.

\begin{figure}[t]
\fbox{
\begin{minipage}{\linewidth}
\internallinenumbers
\setlength{\parskip}{8pt}
\textbf{Box 1: What is a Domain-Specific Programming Language (DSL)?}

When we think of programming languages, we typically think of general-purpose programming languages like Python or C++, which can in principle be used to write programs that perform any computation. But programming language designers have a long tradition of designing small, bespoke programming languages that are tailored for specifying computations that are particular to a given domain \citep{bentley1986programming, bernstein2016perspectives, iverson2007notation}. For example, Stan \citep{carpenter2017stan} is a well-known DSL specialized for Bayesian data analysis. Using Stan, a data scientist can express an analysis plan using specialized syntactic forms like \lstinline{data}, \lstinline{parameters}, and \lstinline{model}. An implementation of this analysis in Stan is likely to be significantly shorter than one in a general-purpose language like Python. It is also likely to run faster, using Stan's built-in efficient inference algorithms. Finally, it is guaranteed to be free of certain bugs, such as incorrectly applying Bayes' rule by forgetting a conditional dependence. However, these benefits of using Stan only apply narrowly within the domain of Bayesian data analysis: for example, a programmer would be ill-advised to implement a video game or a website using Stan. In this way, DSLs restrict the space of programs that can be written, in exchange for making in-domain programs concise, efficient, and correct. Other examples of DSLs include: SQL, a DSL for querying databases (with syntactic forms like \lstinline{SORT} and \lstinline{COUNT}); DOT, a DSL for describing graph diagrams (with syntactic forms like \lstinline{node} and \lstinline{edge}); and CSS, a DSL for specifying typography on web pages (with syntactic forms like \lstinline{font-size} and \lstinline{background-color}).

DSLs can be implemented in many different ways. Some DSLs are implemented from scratch as fresh standalone programming languages. More commonly, however, DSLs are implemented as ``libraries'' of interoperable functions written in a general-purpose programming language \citep{tobin2011languages}. Programming languages designers treat such libraries as languages of their own, because composing the library functions with one another generates a combinatorial space of possible programs that are consistent with the domain theory \citep{hudak1996building}. This paper focuses on memo, which is a standalone language, but our broader argument about intuitive theories holds for languages implemented as libraries, and is consistent with approaches from computational cognitive science that use libraries to represent domain theories \citep{ellis2021dreamcoder, piantadosi2012bootstrapping, grand2024lilo}.
\end{minipage}
}
\end{figure}

Our proposal builds on the influential idea that thinking is like writing and executing programs in a mental ``language of thought'' \citep{fodor1975language, goodman2015concepts, quilty2023best, chater2013programs}. But rather than positing a single ``general-purpose'' language of thought, we posit a collection of domain-specific languag\emph{es} of thought. In our view, our intuitive theories of various domains are encoded in the syntax and semantics of their associated DSLs. To invoke a theory, we write a program in its associated DSL, and to extend that theory, we modify the DSL itself.

To make this idea concrete in the domain of theory of mind, we will examine a real-world DSL called memo \citep{chandra2025domain}. Memo was originally designed as a specialized tool to help cognitive scientists engineer theory-of-mind models in the probabilistic tradition described above. The DSL is built around specialized syntactic constructs like \lstinline{knows}, \lstinline{wants}, and \lstinline{chooses}, which programmers compose to build models of how people reason about particular social situations. These models can then be run efficiently, using GPUs to accelerate the underlying computations. Memo is used by several labs around the world, and memo models have supported analyses reported in several recent papers on social cognition.\footnote{See, for example: \citet{lamba2025solving, collins2025empathy, machino2025minding, chandra2026computational, chen2025signaling, btesh2026action, berke2026toward, chandra2026sycophantic, zur2026pragmatic, chen2026social}.}

In our view, memo is not only a tool for building models of theory-of-mind reasoning in particular scenarios, but also \emph{itself} a model of something more abstract: it is a formal model of $\theta$, the typical human adult's theory-of-mind.
In the rest of this paper, we will develop this idea in three parts, organized around the challenges we identified above.
First, in Section~\ref{sec:structural-variation}, we will review previous approaches to formalizing intuitive theories, and, building on these approaches, we will show how memo solves the problem of formalizing a theory of mind in a way that supports applying that theory to build task-specific models with unbounded structural variation. Second, in Section~\ref{sec:domain-specificity}, we will show how, as a \emph{domain-specific} programming language, memo itself only encodes theory-of-\emph{mind}, but can interface with domain-general world knowledge and intuitive theories of other domains to solve domain-general problems. We will discuss how the mind as a whole might decide when to apply theory-of-mind and which external knowledge to bring to bear, and we will discuss how DSLs can help us understand the value of carving up knowledge into domains in the first place. Third, in Section~\ref{sec:memo-junior}, we will show how varying memo's syntax and semantics allows us to generate alternative programming languages that represent alternative theories of mind. Taking $\theta_3$ as a case study, we will discuss how this perspective can help us study theory of mind development. Finally, by reflecting on the way in which programming languages are themselves programmed, we will suggest a way of applying this same idea recursively to formalize $\Theta$, the space of possible theories of mind, of which $\theta$ and $\theta_3$ are two members.
\section{Formalizing a theory of mind as a DSL}\label{sec:structural-variation}

Why formalize theories of mind as DSLs? Our approach is motivated by a long tradition in computational cognitive science of formalizing intuitive theories in probabilistic terms. (We will discuss connections to a different intellectual tradition, originating in game theory and multi-agent planning, in Section~\ref{sec:i-pomdps}.)

\subsection{From Bayes nets to DSLs}

We take as our starting-point \citet{gopnik2004theory}'s influential proposal to model intuitive theories using the mathematical language of \emph{causal Bayesian networks} or ``Bayes nets'' \citep{pearl2000causality, glymour2001mind}. Bayes nets are appealing representations of intuitive theories for many reasons. First, like theories, Bayes nets specify a quantitative relationship between unobservable abstract entities and observable concrete phenomena. Second, Bayes nets support a number of key theory-like operations such as prediction, planning, inference, and counterfactual reasoning. Third, Bayes nets can themselves be learned from data and experimentation, giving a starting-point for studying conceptual change in the development of intuitive theories.

Could we then represent theories of mind as Bayes nets? We could certainly imagine constructing a Bayes net with nodes for an agent's hidden mental states (e.g.\ beliefs and desires), which are linked to nodes for that agent's observable actions. In fact, this is the approach that \citet{goodman2006intuitive} take in their model of the development of false-belief understanding. This model begins with a simple 4-node Bayes net that represents an agent who acts rationally to achieve its goals based on the true state of the world. After observing data from the agent's behavior, the model learns a more complex 5-node Bayes net that represents an agent who maintains an internal belief state, which is mediated by perception of the true world state, and which affects action. Each Bayes net thus represents a possible theory of mind.

This approach to formalizing theories of mind, though appealing, faces two key challenges. The first problem is that it is not clear which level of abstraction the Bayes net representing a theory of mind should be expressed at. The simple 4/5-node Bayes nets described by \citeauthor{goodman2006intuitive} encode highly abstract, general knowledge of the relationship between beliefs, desires, and actions. But to reason about a particular situation, we need to instantiate this abstract knowledge into a concrete, situation-specific Bayes net, which may be more complex. For example, situations with varying numbers of agents call for Bayes nets with varying numbers of nodes and edges. A single, fixed Bayes net cannot operate at both the abstract and the situation-specific levels of abstraction simultaneously.

Recognizing this limitation of Bayes nets in representing intuitive theories, \citet{tenenbaum2007intuitive} proposed to model intuitive theories using \emph{hierarchical} Bayesian models (HBMs), which link representations of higher-level, more abstract theories to representations of lower-level, more specific theories. Drawing an analogy to linguistics, \citeauthor{tenenbaum2007intuitive} call these higher-level representations \emph{causal grammars:} in the same way that a linguistic grammar generates a constrained space of possible sentences, a causal grammar generates a constrained space of possible Bayes nets. For example, a causal grammar representing a theory of mind might posit a set of abstract node-classes like ``belief-nodes,'' ``desire-nodes,'' and ``action-nodes,'' along with the rule that belief-nodes and desire-nodes influence action-nodes \citep[see worked example in][]{griffiths2007two}. This fixed causal grammar could then be applied to generate situation-specific Bayes nets that vary structurally in the number of nodes and edges. For example, it could generate a six-node Bayes net where one belief and two desires cause an agent to take three actions. Importantly, analogous to \citeauthor{goodman2006intuitive}'s work with Bayes nets, causal grammars too can be learned by Bayesian inference over an even more abstract structure \citep{ullman2012theory, goodman2011learning, kemp2007learning, griffiths2009theory, gopnik2012reconstructing, kemp2008theory, ullman2015nature}. In this way, HBMs solve the problem of abstraction faced by individual Bayes nets.

The second problem with the Bayes net approach---inherited also by HBMs---is that Bayes nets cannot represent recursively-nested inference about inference. This problem is particularly salient in the domain of theory-of-mind, because recursively representing agents who reason about each other is central to social cognition. A single Bayes net can represent ``Alice thinks...'' but to represent a situation where ``Alice thinks Bob thinks...'' we need a more general representation than the individual Bayes net.

One compelling candidate for such a representation is a \emph{probabilistic program} implemented in a probabilistic programming language \citep{goodman2015concepts}. Probabilistic programming languages augment traditional programming languages with constructs for working with random variables, and support for operations like sampling, conditioning, and performing Bayesian inference. Because these constructs compose with each other, and with the rest of the language, a probabilistic program inherits the full expressive richness afforded by a Turing-complete programming language: a probabilistic program can express not only individual Bayes nets, but also more complex structures of recursively-nested and interlinked probabilistic models \citep{goodman2008church}. This makes probabilistic programs better-suited than Bayes nets for representing recursive social reasoning, as demanded by a theory of mind \citep{stuhlmuller2014reasoning}. Indeed, probabilistic programs have long been used for building computational models of recursive social reasoning \citep{zhang2022reasoning, evans2017modeling}.

Probabilistic programs solve the second problem with the Bayes net approach, but they reopen the first problem. We cannot simply associate individual probabilistic programs with theories of mind, because each individual probabilistic program is only a model of a particular situation, and different situations call for different probabilistic programs. The abstract, high-level knowledge that guides a programmer in generating a given probabilistic program to model a given situation remains unspecified.

To address this problem with Bayes nets, \citeauthor{tenenbaum2007intuitive} shifted focus from individual Bayes nets to a causal grammar that generates a space of possible Bayes nets. Could we apply the same approach to probabilistic programs? What we would like is a way to specify a type of ``grammar'' that generates a space, not of possible sentences or Bayes nets, but rather of possible \emph{programs} that are consistent with the theory of a domain. This is exactly what a domain-specialized programming language is: a DSL by definition constrains the space of possible programs to be consistent with the domain theory (see Box~1).

Our proposal, then, is to model a theory of mind as a DSL, which generates probabilistic programs that represent situation-specific models consistent with that theory. Let us now make this proposal concrete with the memo programming language.

\subsection{A brief introduction to the DSL memo}

\begin{figure*}
\centering
\includegraphics[width=\linewidth]{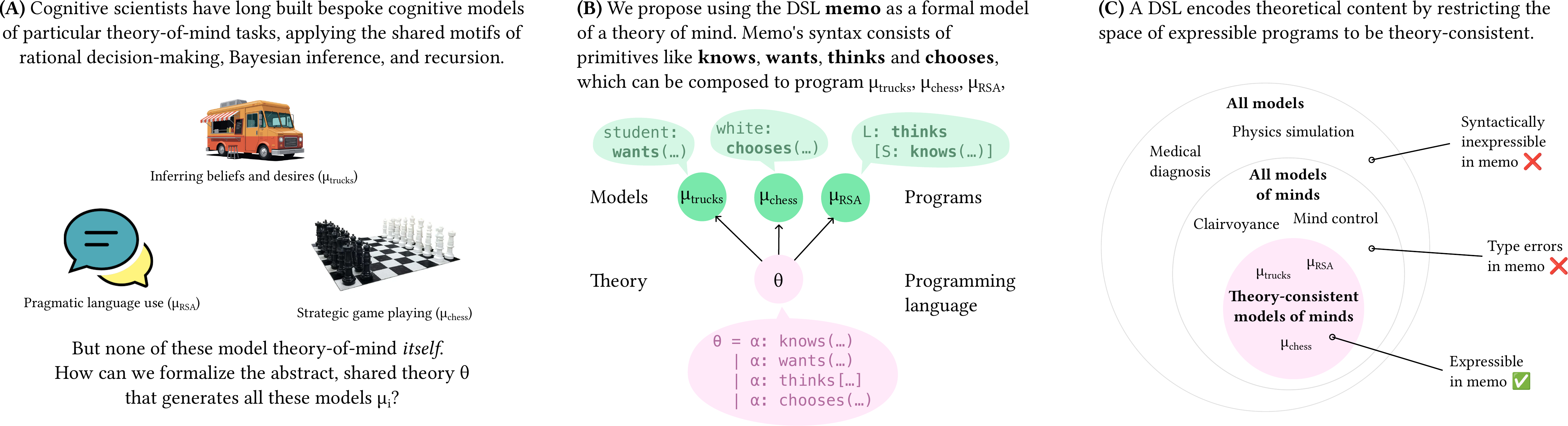}
\caption{How can we formalize the abstract, shared theory of mind $\theta$ that is applied to build various situation-specific models $\mu_i$? In Section~\ref{sec:structural-variation}, we discuss how memo, a domain-specific programming language (DSL) specialized for theory-of-mind reasoning, formalizes a theory of mind. The relationship between a theory and models consistent with that theory is analogous to the relationship between a DSL and programs written in that DSL. Memo's bespoke syntactic forms can be composed to write programs with structural variation in the number of agents and recursive levels of mentalizing.}\label{ref:1-summary-fig}
\end{figure*}

Memo is a domain-specialized probabilistic programming language that is specialized for building models of recursive social reasoning. Memo models are implemented by specifying agents who make a series of choices.

For example, suppose we are modeling an agent named Ali who chooses a bar $b$ to visit after work one day. Suppose that Ali prefers better-quality bars, but does not want to walk far from his workplace. Applying the na\"ive utility calculus \citep{jara2016naive}, we might model Ali as choosing bar $b$ with probability proportional to some function of his total utility, reward minus cost, which in this case is given by \lstinline{quality($b$) - distance($b$)}. In memo, we model such a choice using the \lstinline{chooses} statement: we write that Ali \lstinline{chooses} a bar $b$ from the set of \lstinline{Bars}, with probability proportional (``wpp'') to some function of his total utility. To convert from total utility, which may be negative, to a probability of choosing, which must be positive, computational models typically use the exponentiation function \citep{luce1959individual, franke2023softmax}. Hence, in memo, we write:
\begin{lstlisting}
Ali: chooses(b in Bars, wpp=exp(quality(b) - distance(b)))
\end{lstlisting}
Having set up this model of Ali's choice, we can now use it to make various queries about Ali's choice. For example, we can query the probability that one can find him at a specific bar:
\begin{lstlisting}
return Pr[ Ali.b == {Tony_Tavern} ]
\end{lstlisting}
Or we can compute the expected value of the distance Ali will walk to get to his chosen bar from his workplace:
\begin{lstlisting}
return E[ distance(Ali.b) ]
\end{lstlisting}
In this way, memo allows setting up and querying models of situations involving agents.

So far, we have only used memo to model one agent making one choice. We could have built this model with a single Bayesian network. Unlike a single Bayesian network, however, memo allows us to build models with structural variation. By combining memo's syntactic primitives in different ways, we can use the same, fixed programming language to build models that vary in the number of agents and choices they represent.

For example, suppose Zoe also goes to a bar that evening. Unlike Ali, Zoe is indifferent to distance, but prefers a cheaper location. This means that Zoe's total utility depends, not on \lstinline{distance($b$)}, but on \lstinline{price($b$)}. In memo, we can model Zoe's choice analogously to Ali's:
\newcommand{\exclamation}[0]{!}
\begin{lstlisting}
Ali: chooses(b in Bars, wpp=exp(quality(b) - distance(b)))
Zoe: chooses(b in Bars, wpp=exp(quality(b) - price(b)))
\end{lstlisting}
Our model now contains two agents, Ali and Zoe, who have each chosen bars according to their own personal desires. We can use this model to query the odds that Ali and Zoe will meet:
\begin{lstlisting}
return Pr[Ali.b == Zoe.b]
\end{lstlisting}
Or we can compute how much more we expect Ali to pay than Zoe:
\begin{lstlisting}
return E[price(Ali.b) - price(Zoe.b)]
\end{lstlisting}
Notice that memo separately tracks Ali's choice \lstinline{Ali.b} and Zoe's choice \lstinline{Zoe.b} as two independent random variables. Represented as a Bayesian network, this model would have two nodes, rather than one node as in the previous model of only Ali. In this way, a fixed programming language can allow building models with structural variation in the number of agents and choices.

Another axis of structural variation is in the number of recursively nested levels of thinking present in a model. Suppose that Zoe wishes to meet Ali, but does not know which bar he will go to. To seek out Ali, Zoe might represent her beliefs about how Ali will choose his bar, and then choose her own bar $b$ based on an expected utility that depends on whether she meets Ali.

To represent Zoe's beliefs about Ali's whereabouts, we can use memo's \lstinline{thinks} statement. The \lstinline{thinks} statement is compound: it recursively embeds a memo model that represents Zoe's model of Ali.
\begin{lstlisting}
Zoe: thinks[
  Ali: chooses(b in Bars, wpp=exp(quality(b) - distance(b)))
]
\end{lstlisting}
The \lstinline{thinks} statement can be nested to arbitrary depth: we can write ``Zoe thinks Ali thinks Zoe thinks...'' and so on. For this model, however, we only need the one recursive level implemented above. After having declared Zoe's beliefs about Ali, we can model Zoe's choice of bar based on her mental state. Let us say that Zoe's utility is 10 if she meets Ali, and 0 otherwise. We can write her expected utility as \lstinline{E[10 if b == Ali.b else 0]}. Hence, we can model her choice as:
\begin{lstlisting}
Zoe: chooses(b in Bars, wpp=exp(E[10 if b == Ali.b else 0]))
\end{lstlisting}
Let us make two observations about this line of code. First, to be clear, the \lstinline{Ali.b} in this expression refers to Zoe's belief about Ali's location, not Ali's actual location. The memo compiler keeps these entities separate: for example, if Zoe had a false belief about Ali's location, then her choice would be taken with respect to her false belief, not with respect to Ali's true location. We will discuss this point further in Section~\ref{sec:memo-junior}. Second, because Zoe has uncertainty about Ali's choice, we modeled her as computing an \emph{expected} utility using the \lstinline{E[...]} expression to marginalize over her uncertainty. We will return to this point later in this section.% Section~\ref{sec:theoretical-content}.

We can use this simple model of Zoe's choice to make a variety of predictions. For example, we can predict the likelihood that Zoe will choose to search for Ali at Tony's Tavern:
\begin{lstlisting}
return Pr[ Zoe.b == {Tony_Tavern} ]
\end{lstlisting}
Or we can predict how likely Zoe thinks it is that she will meet Ali:
\begin{lstlisting}
return E[ Zoe[ Pr[b == Ali.b] ] ]
\end{lstlisting}
Here, we use the expression \lstinline{Zoe[...]} to compute the expression \lstinline{Pr[b == Ali.b]} from Zoe's perspective.

As a last example, suppose that Zoe doesn't know where Ali works---perhaps she has a uniform prior belief over a set of possible \lstinline{Jobs} that Ali could have. We can amend her mental model of Ali to account for this additional layer of uncertainty.
\begin{lstlisting}
Zoe: thinks[
  Ali: chooses(j in Jobs, wpp=1),  # uniform prior
  Ali: chooses(b in Bars, wpp=exp(quality(b) - distance(j, b)))
]
\end{lstlisting}
If Zoe does succeed in meeting Ali at the bar, she can apply Bayes' rule to update her belief about Ali's job. For example, she might conclude that Ali is likelier to work at an office that is near that bar. To model this inference, we use memo's \lstinline{observes} statement, which updates the observer's belief state to be consistent with the observed evidence.
\begin{lstlisting}
Zoe: observes Ali.b is Zoe.b  # they meet
\end{lstlisting}
Now, we can predict Zoe's posterior belief, after meeting Ali, that he works at City Hall:
\begin{lstlisting}	
return E[ Zoe[ Pr[Ali.j == {City_Hall} ] ] ]
\end{lstlisting}
Or, we can imagine that on the next day, Ali chooses a new bar $b^\prime$. If Zoe wants to meet Ali again, how might she change her strategy based on the information about $j$ she gleaned the day before? We might expect her to choose a bar near the inferred $j$.
\begin{lstlisting}
Zoe: thinks[
  Ali: chooses(b$^\prime$ in Bars, wpp=exp(quality(b$^\prime$) - distance(j, b$^\prime$)))
]
Zoe: chooses(b$^\prime$ in Bars, wpp=exp(E[10 if b$^\prime$ == Ali.b$^\prime$ else 0]))
\end{lstlisting}
This more complex model is impossible to represent as a single Bayesian network because it requires nested inference: Zoe's choice of $b^\prime$ depends on her inference of Ali's $j$.

Let us summarize our progress so far. Over the course of the preceding paragraphs, we wrote memo models of three different situations: Ali alone, Zoe-and-Ali, and Zoe-seeking-Ali. These situation-specific models $\mu_{1,2,3}$ varied in the number of agents, choices, and recursive levels of reasoning they represented. However, they were all expressed in the same fixed programming language memo. In this way, we propose that memo itself is a model of the shared theory-of-mind $\theta$ that was applied to generate the situation-specific models $\mu_{1,2,3}$.

\subsection{What theoretical content is encoded in memo?}\label{sec:theoretical-content}

All of the models in the previous section can in principle be implemented without using memo. The computations can be implemented either by hand, by using any general-purpose programming language (like Python or C++), or by using a probabilistic programming language (like Church or WebPPL) that automates the computations necessary for Bayesian inference. In fact, memo statements like \lstinline{chooses}, \lstinline{observes}, and \lstinline{thinks} correspond roughly to the operations of \emph{sampling}, \emph{conditioning}, and \emph{recursive querying} in a probabilistic programming language, allowing for a nearly one-to-one translation between memo models and Church or WebPPL programs. More generally, by the logic of Turing-completeness, any memo model must be expressible in any general-purpose programming language like Python or C++.

Why, then, do we say that memo models $\theta$ but Python, C++, Church, and WebPPL don't? In what way does memo's design encode the content of a theory of mind? The answer to this question lies, not in what memo can model, but in what memo cannot model. Memo's design explicitly prevents programmers from writing models that are inconsistent with theory $\theta$.

For example, recall that in our model of Zoe-seeking-Ali, we had to compute Zoe's \emph{expected} utility of finding Ali, because Zoe was uncertain about Ali's location. Suppose that we accidentally omitted the expectation:
\begin{lstlisting}
Zoe: thinks[
  Ali: chooses(b in Bars, wpp=exp(quality(b) - distance(b)))
]
Zoe: chooses(b in Bars, wpp=exp(10 if b == Ali.b else 0))
\end{lstlisting}
The problem with this model is that it is inconsistent with our theory of mind. If Zoe is uncertain about Ali's choice of bar, then it does not make sense for her to simply choose to go to the bar that he is in, \lstinline{Ali.b}. That would be clairvoyance. Indeed, memo detects this issue and refuses to run this model, raising an error message for the programmer.

An ordinary probabilistic programming language, however, will gladly admit and execute this theory-inconsistent model. Consider how we might implement Zoe-seeks-Ali in the probabilistic programming language WebPPL. We will present an implementation lightly adapted from a textbook by \citet{evans2017modeling}. In WebPPL, Ali's choice of a nearby bar is traditionally implemented by first using the \lstinline[language=webppl]{uniformDraw} construct to declare that Ali chooses a bar, and then using the \lstinline[language=webppl]{factor} construct to bias the inference engine to sample high-utility bars:
\begin{lstlisting}[language=webppl]
ali_b = uniformDraw(Bars);
factor(quality(ali_b) - distance(ali_b));
return ali_b;
\end{lstlisting}
This model produces the correct answer. But what about Zoe's choice of a bar with the goal of finding Ali? Reasoning by analogy to the previous model, it is tempting to write the following:
\begin{lstlisting}[language=webppl]
ali_b = uniformDraw(Bars);
factor(quality(ali_b) - distance(ali_b));

zoe_b = uniformDraw(Bars);
factor(10 if zoe_b == ali_b else 0);

return zoe_b;
\end{lstlisting}
WebPPL will gladly execute this model, but the model produces the wrong numerical answer. We can see the error most clearly if, instead of predicting Zoe's choice of bar, we predict the probability that they meet:
\begin{lstlisting}
return zoe_b == ali_b;
\end{lstlisting}
The model erroneously predicts a near-100\% chance that they meet, even though there is no way for Zoe to guarantee meeting Ali if she is uncertain of Ali's location.

The problem is that WebPPL has no language-level support for the notion of associating choices with particular agents. WebPPL deals only in random variables. Zoe's choice of \lstinline{zoe_b}, and Zoe's belief about Ali's choice of \lstinline{ali_b}, are indistinguishable at the language level, with no way of representing that Zoe chooses and knows \lstinline{zoe_b} but has no control over and is uncertain about \lstinline{ali_b}. This makes it easy to accidentally mix these variables in a way that is inconsistent with our theory of mind. A \lstinline[language=webppl]{factor} statement intended to affect an action choice might instead affect a belief, or vice versa---for example, we would run into a similar bug if we tried to model Zoe learning about Ali's location by observing that they met at the bar. It is quite common for even experienced probabilistic programmers to make this type of mistake in practice \citep{levine2018reinforcement}.

In contrast, memo provides explicit syntax to distinguish these two types of variables via the \lstinline{thinks} and \lstinline{chooses} statements, which are syntactically marked by the agent doing the thinking or choosing. Furthermore, memo's semantics explicitly detect and forbid conceptually incoherent models before those models are even run. \citet[\S3.1]{chandra2025domain} discuss a variety of ``principles of agency'' that are formally encoded in memo's compiler as rules for when to accept or reject programs. Each of these rules represents theoretical content, formalizing some aspect of our theory of mind.

In summary, while general-purpose programming languages give us the building-blocks to express a vast range of computations that accord with our theory-of-mind, those same building-blocks can also be assembled to express computations that are inconsistent with our theory of mind. Memo, in contrast, provides syntax and semantics that constrain the space of admissible models to those consistent with our theory of mind. In this way, memo's syntax and semantics formally encode the content of a theory of mind.

\subsection{Connections to game theory and multi-agent planning}\label{sec:i-pomdps}

We began this section by working forwards from influential proposals for formal representations of intuitive theories---Bayes nets and causal grammars---and describing how the DSL is a natural extension that supports formalizing a theory of mind. But an alternate approach to formalizing theories of mind might take as its \emph{starting-point} the recursive, multi-agent nature of theories-of-mind. In particular, decades of work in game theory, multi-agent planning, and reinforcement learning has developed formalisms for representing multi-agent scenarios: Markov Games \citep[MGs;][]{shapley1953stochastic, littman1994markov} and Partially Observable Markov Games \citep[POMGs;][]{hansen2004dynamic, kuhn1953extensive} for competitive scenarios, Decentralized Partially Observable Markov Decision Processes \citep[Dec-POMDPs;][]{bernstein2002complexity} and Network Distributed Partially Observable Markov Decision Processes \citep[ND-POMDPs;]{nair2005networked} for cooperative scenarios, and Interactive Partially Observable Markov Decision Processes \citep[I-POMDPs;][]{gmytrasiewicz2005framework} for scenarios requiring recursive inference over other agents' beliefs. More recently, \citet{barnby2024beyond} have offered a framework called the Standard Framework for Social Cognition (SFSC).

These formalisms are theory-like in at least two ways. First, like causal grammars and DSLs, they exist at a higher level of abstraction than situation-specific models do. The formalisms define a set of abstract parameters, such as ``state spaces,'' ``action spaces,'' and ``reward functions,'' as well as rules for deriving the optimal policies of rational agents in a situation governed by those parameters. To apply a formalism to model a particular situation, we can instantiate that formalism with concrete parameter values: for example, by providing a specific state space, action space, and reward function. Second, as we described with memo, the design of these formalisms holds theoretical content. For example, \citeauthor{gmytrasiewicz2005framework} describe how the assumptions that agents cannot read or control each other's minds are formally encoded into the types of the observation and transition functions specified by the I-POMDP.

Could we say then that MGs, POMGs, Dec-POMDPs, ND-POMDPs, I-POMDPs, and the SFSC are an alternative formalization of theories of mind?
%There are two challenges with this proposal. First, the formalisms of game theory and reinforcement learning were designed to support planning and prediction via the computation of optimal policies. Unlike Bayes nets, these formalisms require additional machinery to support other theory-like operations, such as inference and counterfactual reasoning. For example, \citet{ng2000algorithms} extend reinforcement learning formalisms with additional machinery to infer an agent's reward function. In contrast, this inference is natively expressible in memo with no additional machinery.
Rather than seeing these formalisms as competing alternatives, we see them as consistent with and generalized by DSLs. Let us explain. If we were to adopt one of these formalisms to represent a theory of mind, we would be faced with the problem of choosing which one to use. But the best-suited formalism varies by scenario: for example, many scenarios are well-modeled with Dec-POMDPs, while some require the more complex machinery of I-POMDPs. A natural choice might be to select the most general formalism, the I-POMDP, as our representative, but this would run up against a computational problem: precisely because the I-POMDP is fully general, computing optimal policies in an I-POMDP is often intractable.\footnote{Solving an I-POMDP is in the computational complexity class PSPACE-hard, so ``at least as hard'' as the hardest NP-complete problem \citep{gmytrasiewicz2005framework}. See \citet[figure 27.2]{kochenderfer2022algorithms} for a detailed taxonomy of these formalisms, their relative generality, and their complexity classes.} Even though it is technically possible to encode all scenarios as I-POMDPs, by doing so we sacrifice computational tractability for a large class of simple scenarios that could have been expressed in terms of the lighter-weight Dec-POMDP formalism.

Instead of being forced to choose a single formalism, what we would like is a unifying meta-formalism that allows us to flexibly construct exactly the features needed for modeling a given scenario, trading off expressivity and computational efficiency as called for by the scenario. In fact, memo is exactly such a meta-formalism: memo can be used to implement all of the formalisms listed above, from MGs to I-POMDPs,\footnote{For example, consider this memo implementation of the most complex formalism, the I-POMDP, instantiated to model a player in \citet{berg1995trust}'s Investment Game: \url{https://github.com/kach/memo/blob/main/demo/demo-i-pomdp.ipynb}.} as well as to mix and match features to create bespoke new formalisms that have never been expressed before. For example, we can use memo to express a situation where two agents are engaged in a competitive MG, and three others are independently engaged in a collaborative Dec-POMDP. Furthermore, without any additional machinery we can then recursively embed this entire model inside a model of an observer who is inferring those five agents' mental states.

In this way, DSLs bridge the strengths of the formalisms from game theory and reinforcement learning (native support for recursive social reasoning) with the strengths of the Bayes net and causal grammar tradition (native support for theory-like operations like inference, counterfactual reasoning, causal learning). By attaining the best of both worlds, DSLs not only address the first challenge of formalizing the shared theory-of-mind behind all situation-specific models, but also pave the way for addressing the second challenge of explaining how theory-of-mind interacts with other knowledge, and the third challenge of formalizing alternate theories of mind. It is to these challenges that we turn in the rest of this paper.

\section{Domain-specificity in minds and machines}\label{sec:domain-specificity}

When we set ourselves the goal of formalizing theories of mind, we stipulated that there is such a thing as a ``theory of mind'' at all; that is, that humans have a well-defined domain-specialized system for reasoning about others' mental states, and that the knowledge applied by that domain-specialized system can be isolated from the rest of the knowledge possessed by the mind.
A wealth of empirical evidence---behavioral, developmental, neural---supports this view (see \citet{hirschfeld1994mapping, kanwisher2010functional, carey2009origin, spelke2022babies, cosmides1997evolutionary}; and more recently a review by \citet{liu2025physical}).
But all of this evidence faces a key theoretical challenge, famously raised by \citet[p.~103]{fodor1983modularity}: \emph{How can there be any modular, domain-specialized cognitive systems at all?}

\citeauthor{fodor1983modularity}'s concern is that the essence of modularity is the encapsulation of information: modules should have limited access to external knowledge. But on \citeauthor{fodor1983modularity}'s view, information encapsulation is incompatible with the work of cognitive systems. Cognitive systems are responsible for belief fixation, and belief fixation by its very nature requires unfettered global access to information, because any bit of world knowledge could potentially be relevant to the problem at hand. ``In principle, our botany constrains our astronomy,'' writes \citeauthor{fodor1983modularity}---or, specializing this argument to the domain of social cognition, \citet{currie2000think} write: ``Just where are the bits of information to which we are systematically blind in making our social judgements? The whole genre of the detective story depends on our interest and skill in bringing improbable bits of far-away information to bear on the question of someone's credibility.''\footnote{See also \citet{garfield2001social}, who write, ``ToM reasoning is simply too dependent on general knowledge about the goals, attitudes and information available to those about whom we are reasoning, and it is, well, too much a matter of reasoning to be encapsulated.''}
A domain-specialized system is not useful without a mechanism for applying that system to reason about arbitrary entities in the open world. But it is not clear how to design a system that encapsulates its domain-specific knowledge while operating over general world knowledge---or whether such a system, if built, would end up counting as domain-specific at all. This brings us to the second challenge we raised in the introduction to this paper: Is it possible to formalize \emph{only} theory-of-mind, no more, no less?

The difficulty with formalizing ``only theory-of-mind'' is reflected in traditional computational models\footnote{Here, we are concerned specifically with the challenge of building computational models that reflect modular organization of \emph{knowledge} in the mind. The possible modularity of perception, motor control, and memory, which have long been reflected in cognitive architectures like ACT-R \citep{anderson2009can} and Soar \citep{laird1987soar}, are complementary axes of modularity in the mind.} of theory-of-mind reasoning. Consider for example \citet{baker2017rational}'s Bayesian inverse planning model of belief and desire attribution in the ``food trucks'' paradigm. This model takes as input a cartoon video of a hungry graduate student searching for a food truck, and produces judgments of the graduate student's preferences over food truck cuisine options, as well as the graduate student's initial (possibly-false) beliefs about where various food trucks are parked. The outputs of this model predict human judgments very well, suggesting that the model captures important aspects of how people reason about these scenarios.

But, taking a step back, what exactly is \citeauthor{baker2017rational}'s model a model \emph{of?} In principle, it is a model of theory-of-mind reasoning, and indeed, the model's source code includes procedures for performing computations that should clearly be considered part of theory-of-mind under the Bayesian inverse planning framework (e.g.\ procedures for computing noisily-rational actions and for performing Bayesian belief updates). But the source code to this model is also peppered with procedures for a wide variety of \emph{other} computations, computations that seem to have nothing to do with theory-of-mind per se: for example, procedures for computing the walking distance between two parking spots, or reasoning about the impenetrability of a brick wall. These food-truck-and-campus-layout-specific procedures are interleaved with the source code of the full Bayesian inverse planning model, tucked between procedures for computing noisily-rational action and Bayesian belief updates. This hardly seems like a model of a modular system: there is no clear boundary that demarcates where the theory of mind ends and the theory of parking spots begins. In this way, \citeauthor{baker2017rational}'s model excels as a model of how people reason specifically about graduate students foraging at food trucks, but it is not a model of theory-of-mind itself.

A Fodorean skeptic might say that this is no surprise: in fact, it is impossible to deliver such a model. Any effort to build a computational model of theory-of-mind itself is condemned to fail in one of two ways. One failure mode is for the model to be so pure as to be unable to account for any concrete real-world phenomena. After all, without building in some notion of distances and walls into the food trucks model, it would be impossible to reason about the relative costs of the graduate student's potential routes, and thus impossible to apply the na\"ive utility calculus to make mental state inferences. The other failure mode is including so much non-theory-of-mind-related world knowledge as to defeat any plausible claims of modularity.

How can we reconcile the empirical evidence in favor of a domain-specialized system with the theoretical challenge raised by \citeauthor{fodor1983modularity}? Our proposal, to formalize theories-of-mind not as programs but as programming \emph{languages}, allows us to make progress here. In fact, ``domain-specificity,'' ``modularity,'' and ``encapsulation'' are all common terms in the programming languages literature \citep[see, e.g.,][for some examples of seminal papers on these topics]{parnas1972criteria, boyapati2003ownership, bentley1986programming, leroy1994manifest, hudak1998modular}. Modular software architectures have many pragmatic benefits: modules can be reused across projects, they can be checked for correctness in isolation, they can be developed independently and in parallel by different teams, and they can be used by non-specialists who are agnostic to the technical implementation of the module's internals. Every DSL designer faces the challenge of designing a system that cleanly abstracts away the particulars of its respective domain, while still being useful in the real world. Examining how DSL designers navigate this tension can help us understand how domain-specific cognitive systems may be architected to participate in domain-general computations.

\subsection{Accessing domain-general knowledge in domain-specific languages}

To study how DSL designers navigate this tension, let us consider as a case study the Structured Query Language (SQL), a widely-used DSL that is specialized for making queries to databases. SQL's core competence is in helping programmers perform database search tasks like ``search this database of real estate listings for the 10 cheapest 2-bedroom apartments within a mile of the nearest train station'' or ``search this database of student records for students enrolled in Calculus without the prerequisite of Algebra.'' In the same way that memo is a formalization of a theory of mind, which is used to efficiently represent agents' mental states and predict their behavior, SQL is a formalization of a theory of structured relational data \citep{codd1970relational}, which is used to efficiently represent database schemas and execute queries over them. Just as memo has built-in algorithms for performing Bayesian inference about hidden mental states, SQL's compiler has built-in algorithms for efficiently sorting, indexing, and filtering rows of databases.

To answer a query about 2-bedroom apartments, a SQL program must be able to reason over concepts of apartments and bedrooms---at the very least, to understand that every apartment is associated with some number of bedrooms. This knowledge is clearly beyond the scope of the general theory of structured relational data, and a domain-specific language for database queries should not be endowed with bespoke knowledge of bedrooms (nor should a domain-specific system for reasoning about other agents be endowed with bespoke knowledge of food trucks). SQL must thus be designed so that without any such built-in knowledge, it is able to answer queries about apartments and bedrooms (not to mention employees and managers, books and authors, courses and instructors, and all other facets of the ever-changing real world). How is this possible? SQL accomplishes this by requiring programmers to provide the relevant world knowledge themselves. Programmers who use SQL are responsible for defining the ontology of relevant data types and relations when configuring their database schemas, as well as for defining helper functions (e.g.\ for computing the walking distance between an apartment and the nearest train station) when performing queries. Equipped with this bespoke, curated slice of world knowledge, SQL can then draw on its core competence in processing structured relational data to efficiently represent and execute the necessary database queries. Importantly, SQL itself---that is, its compiler, which contains the implementations of these efficient database operations---does not include any information about apartments or train stations.

This example shows that domain-specific systems can be designed to flexibly accept incoming exogenous world knowledge on an ad-hoc, situation-specific basis. This lesson echoes theoretical work in cognitive science on the issue of modularity and encapsulation. For example, responding to \citeauthor{currie2000think}'s example of the detective story, \citet{carruthers2006architecture} argues  that ``All that the example really shows is that the mind‐reading faculty may need to work in conjunction with other elements of cognition in providing us with a solution to a problem, querying other systems for information. In fact,'' he continues, ``most of the burden in detective-work is placed on physical enquiries of one sort or another---investigating foot‐prints, finger‐prints, and so forth.'' \citeauthor{carruthers2006architecture} advocates for distinguishing between ``narrow-scope'' and ``wide-scope'' notions of encapsulation. In both cases, the encapsulated system has native internal information specific to its domain. But they differ in the degree to which they can access external information. A narrowly encapsulated system can only access a small, \emph{determinate} subset of the available external information, while a widely encapsulated system can be dynamically configured to query \emph{any} small subset of the total available external information. In principle, wide-scope encapsulation retains the key computational benefit of narrow-scope encapsulation---namely, what \citeauthor{carruthers2006architecture} calls ``computational frugality,'' in the sense of limited use of computational resources---but in addition, it makes it possible for cognitive systems to be modular after all.

Our proposal is that DSLs provide a concrete computational account of cognitive systems with wide-scope encapsulation, and thus of systems that are domain-specialized while being able to work with general world knowledge. Let us return to the domain of social cognition, as modeled by the DSL memo, for an example. In the same way that SQL can be used to query databases of apartment listings without itself having any knowledge of apartments built-in, we will show that memo can be used to reason about social situations involving arbitrary external world knowledge, without itself having any of that knowledge built-in.

How would we implement \citeauthor{baker2017rational}'s food trucks model in memo? The full model is available online\footnote{\url{https://osf.io/x4uak/overview?view_only=01eaf2089bbf4eb1982b2b764977f798}}, but let us focus on a couple of key highlights here. First, the model sets up some basic data types that are necessary to reason over this scenario, such as the space $S$ of possible locations on a 2D grid, and the space $A$ of possible actions that the hungry grad student can take at any given moment. To be clear, this is \emph{not} memo code, but rather ordinary Python code that we will reference in memo shortly.
\begin{lstlisting}
S = pair(x=range(0, WIDTH), y=range(0, HEIGHT))
class A(Enum):
  LEFT = 0; RIGHT = 1; UP = 2; DOWN = 3
\end{lstlisting}
A helper function then describes how taking an action $a \in A$ changes the grad student's location from $s \in S$ via vector addition. (The full code includes a little more logic to account for the impenetrability of walls.)
\begin{lstlisting}
def take_a_step(s, a):
    deltas = array([[-1, 0], [+1, 0], [0, -1], [0, +1]])
    return S(s + deltas[a])
\end{lstlisting}
Having defined this basic ``general world knowledge,'' we can now use memo to model the grad student taking a step in the direction that maximizes the chances that (for example) she ends up in a parking spot where there is food.
\begin{lstlisting}
student: knows(s)
student: wants(goal=food_at_state($s^\prime$))
student: chooses(a in A, to_maximize=EU[goal])
student: given($s^\prime$ in S, wpp=$s^\prime$ == take_a_step(s, a))
\end{lstlisting}
Notice that the memo code is agnostic to how $S$, $A$, and \lstinline$take_a_step$ are defined, or what they do. That general world knowledge was expressed in Python, a general-purpose programming language. memo is however able to accept this incoming exogenous information and integrate it into the model in order to perform theory-of-mind-specific reasoning over this world knowledge.

The Fodorean skeptic might still object that we have not really made any progress with this model. After all, isn't this model ultimately just like \citeauthor{baker2017rational}'s, in the sense that code modeling parking-lot-geometry is juxtaposed with code modeling rational action? And doesn't the memo code still need to know at least about the \emph{existence} of $S$, $A$, and \lstinline$take_a_step$---which means it is not truly a modular system after all? Here, it is important to remember that this code is \emph{not} meant to be a model of theory-of-mind. This code is indeed a model of how people reason about hungry grad students foraging at food trucks. The theory-of-mind itself is modeled by code that is not even pictured here: namely, the code that implements memo's \emph{compiler}. The compiler is where the implementations of noisily-rational action and Bayesian belief updating lie: it is shared unmodified among all memo users and all memo models, and it is agnostic to even the existence of $S$, $A$, and \lstinline$take_a_step$. The memo compiler knows how to work with agents, action spaces, and utility functions (concepts that are posited as part of the theory of mind), but it has no built-in knowledge of any \emph{particular} agents, action spaces, or utility functions. Those must be provided by memo programmers. In this way, the memo compiler is a domain-specialized and informationally-encapsulated system that is nonetheless able to interface with general world knowledge.

\subsection{Composing multiple domain-specialized systems}

\begin{figure*}
\centering
\includegraphics[width=\linewidth]{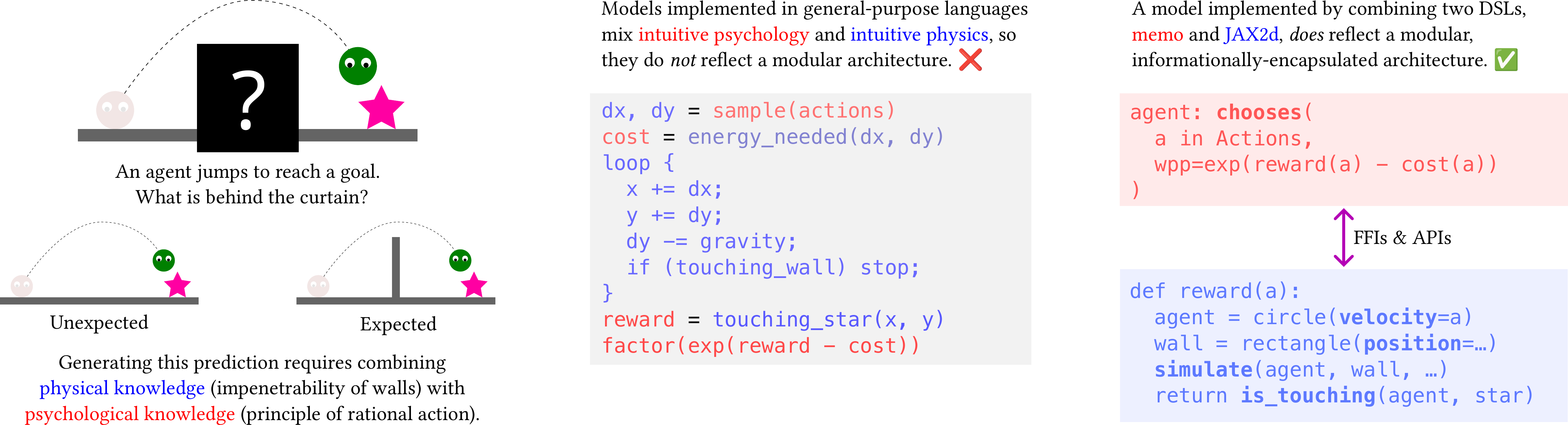}
\caption{How can a theory of mind be domain-specialized and informationally encapsulated, yet interface with external world knowledge? In Section~\ref{sec:domain-specificity}, we discuss how treating theories of minds as DSLs helps answer this puzzle: multiple DSLs can be composed to build models that combine knowledge from multiple intuitive theories. In this example, we combine intuitive physics and intuitive psychology to reason about a classic stimulus by \citet{csibra2003one}.}\label{ref:2-summary-fig}
\end{figure*}

In our discussion so far, we have used facts like ``the impenetrability of walls'' as examples of exogenous world knowledge that is outside of theory-of-mind. But the impenetrability of solid objects is not an isolated fact about the world---rather it itself is a key principle of a different intuitive theory: an intuitive theory of physics. Let us turn, then, to the question of how a domain-specialized cognitive system could interface, not just with isolated bits of exogenous knowledge, but with another domain-specialized cognitive system.

Consider the case of intuitive psychology and intuitive physics. As \citet{liu2025physical} discuss, there is substantial evidence that people have separate domain-specialized systems for reasoning about the mental states of agents in the social world, and for reasoning about the dynamics of objects in the physical world. But in most real-world situations involving other people, we must simultaneously reason about both of these types of entities in a deeply integrated manner. \citeauthor{liu2025physical} point out that this problem arises even in the simplest and earliest examples of such reasoning, such as in \citet{csibra2003one}'s classic experiments with one-year-old infants. Following \citeauthor{liu2025physical}, we will take \citeauthor{csibra2003one}'s simple and elegant stimuli as a running example in this section, though our claims are about the representations held by adults, not infants.

\citet{csibra2003one}'s stimuli show an agent jumping to reach a target object. The space above which the agent jumps is hidden to the observer, but observers infer that this space likely contains a wall or barrier over which the agent jumped. To make this inference, the observer must reason both about the agent's physical body (e.g.\ the effortfulness of jumping, the impenetrability of solid barriers) and the agent's mental states (e.g.\ the principle of rational action \citep{jara2016naive}, by which the agent is unlikely to take an effortful and costly jumping action in the absence of a barrier). How do intuitive physics and intuitive psychology interface with one another to solve this problem?

To approach this issue from a DSL perspective, let us treat these two different domain-specialized cognitive systems (intuitive physics and intuitive psychology) as two different DSLs. Thus far, we have focused on treating theory-of-mind as a memo-like DSL, but implicit in our proposal has been the idea that \emph{any} intuitive theory could be thought of as a DSL. Indeed, a growing body of work argues that rather than thinking about ``the'' language of thought, we should think about the many different domain-specific languag\emph{es} of thought present in a given mind (\citealp{mandelbaum2022problems, dehaene2022symbols, quilty2023best}; \citealp[\S6.3]{carruthers2006architecture}). The history of programming languages is full of examples of DSLs that formalize theories of all kinds of domains, ranging from the theories of scientific domains like classical mechanics \citep{sussman2015structure} to ray optics \citep{hanrahan1990language}, color perception \citep{chen2024coolerspace}, and differential geometry \citep{li2024mesh}, to the theories of culturally-constructed activities like knitting \citep{lin2023semantics} and music \citep{anders2011constraint}.

If we adopt this perspective, then the question becomes: {Is it possible to write ``multilingual'' \citep{felleisen2018programmable} programs that combine multiple domain-specific languages to solve domain-general problems?} Programming language theory tells us that the answer is yes. For example, to build a website, a programmer might use the language PHP to maintain the web server, the language SQL to query the database, the language HTML to specify the page layout, the language CSS to design the fonts and colors, and the language JavaScript to implement the behaviors of interactive buttons. These five languages are specialized to their various domains, but they work together in concert. When a visitor to a website clicks on a button, a bit of JavaScript code is activated, which interfaces with a bit of PHP code on the web server, which interfaces with some SQL code in order to query the database, and finally the result is passed back from SQL to PHP to JavaScript and displayed to the visitor by interfacing with the HTML and CSS. How is this multilingual interaction implemented? Programming languages talk to one another using two paired mechanisms. A ``foreign function interface'' (FFI) allows code in programming language $A$ to make outgoing requests to run code written in programming language $B$, while an ``application programming interface'' (API) allows code written in programming language $B$ to accept incoming requests to run code from programming language $A$. For example, to query a database from a web server, a programmer might connect PHP's FFI to SQL's API.

Let us see how FFIs and APIs can be used to build a model that reasons about an agent jumping to reach a target object, as in \citeauthor{csibra2003one}'s stimuli, by combining memo with a different DSL that captures intuitive physics.

Precisely characterizing this putative language-of-thought-of-physics is beyond the scope of this paper, but let us briefly describe what such a language might look like. One way intuitive physics has been characterized in computational terms is by analogy to a ``video game engine,'' a system that supports setting up and efficiently simulating ``levels'' of a game (i.e.\ physical scenarios) based on resource-rational approximations to the laws of physics \citep{ullman2017mind, battaglia2013simulation}. Perhaps unsurprisingly, video game designers have long developed DSLs that make it easier to set up games and levels. These languages have evolved with advancements in computer graphics, but almost always revolve around specifying objects and their causal relationships. Very early text-based games like \emph{Zork} were implemented using languages like Zork Implementation Language (ZIL) \citep{meretzky1989learning} and Inform \citep{nelson1993inform}, which provide syntax for declaring objects, their locations and spatial relationships (e.g.\ support or containment), and abstract rules for how their properties change in response to interactions with other objects. Primitive graphical games, including Atari-like games like \emph{Frogger}, can be specified using languages like Video Game Description Language (VGDL) \citep{ebner2013towards, schaul2013video} or PuzzleScript \citep{lavelle2013puzzlescript}, which provide syntax for specifying the spatial positions of objects on a 2D grid, as well as how these objects move and interact upon contact with one another. Finally, modern video games with rich, continuous-space rigid-body dynamics, like \emph{Angry Birds}, are implemented using frameworks like box2d \citep{catto2007box2d}, which provide syntax for specifying objects and their positions, velocities, and physical properties, as well as tools for simulating Newtonian mechanics on these scenes. In the same way that the syntax and semantics of memo formalize psychological concepts like agents, beliefs, and desires, the syntax and semantics of languages like ZIL, Inform, VGDL, PuzzleScript, and box2d formalize physical concepts like bodies, positions, contact, containment, forces, and torques. In this way, we might take these languages to be formalizations of intuitive theories of physics.

\begin{figure*}
\centering
\includegraphics[width=\linewidth]{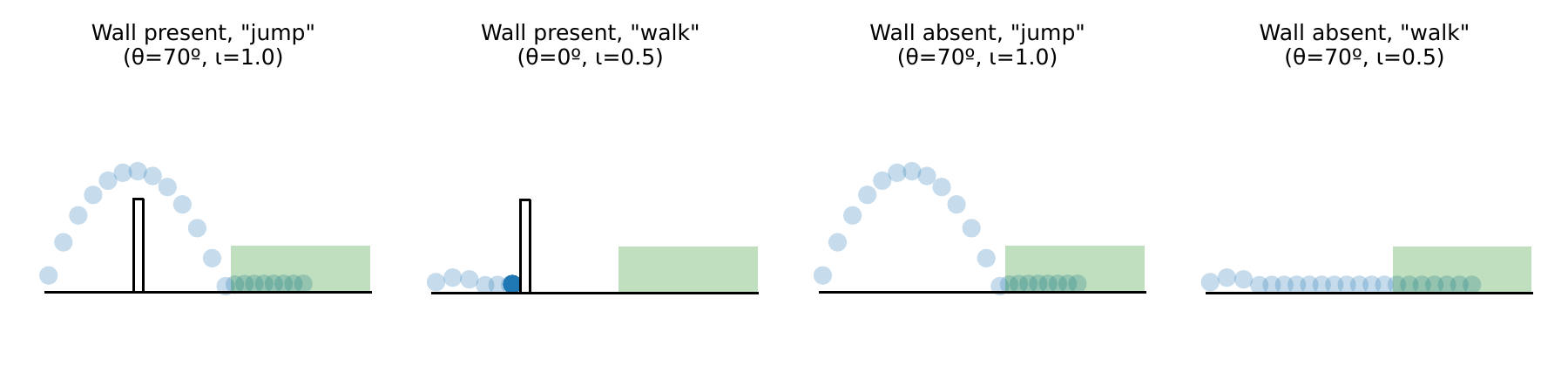}
\caption{Sample simulations from our JAX2d simulator, varying the impulse $\iota$ and the angle $\theta$ to produce ``jump'' and ``walk'' motions, in the presence and absence of an immovable wall.}\label{fig:sim-demo}
\end{figure*}

To model people's intuitions about \citeauthor{csibra2003one}'s stimuli, we will use a physics engine called JAX2d \citep{matthews2024kinetix}, which is a modern, lightweight variant of box2d. JAX2d has an API that lets us manipulate JAX2d scenes using Python code. For example, in JAX2d, we can set up a scene with an agent who has a circular body like this (code slightly simplified for readability):
\begin{lstlisting}
scene = create_empty_sim()
add_circle_to_scene(
  scene, radius=0.1, position=[0, 0.1], velocity=[0, 0],
  friction=0, density=1
)
\end{lstlisting}
Similarly, we can conditionally install an immovable wall by adding a rectangular body:
\begin{lstlisting}
if wall:
  add_rectangle_to_scene(
    scene,
    position=[1., 0.5],
    dimensions=[0.1, 1.0],
    fixated=True  # immovable
  )
\end{lstlisting}
JAX2d's API exposes a Python function \lstinline{step(scene)}, which steps the simulation of \lstinline{scene} forward one tick in time. We can optionally supply external forces that act on each body in the scene. For example, we can write \lstinline{step(sim_state, [0.0, 1.0])} to give a vertical kick to the agent, i.e.\ an impulse of 1.0 units in the $+Y$ direction. Putting this together with our scene setup code, we can write a simple function that simulates an agent moving by giving it an impulse $\iota$ in direction $\theta$ at time $t=0$, and simulating for 100 more steps.
\begin{lstlisting}
def simulate(wall, $\iota$, $\theta$):
  scene = create_empty_sim()
  agent = add_circle_to_scene(...)
  if wall: add_rectangle_to_scene(...)
  scene = step(scene, [$\iota \cdot \cos(\theta)$, $\iota \cdot \sin(\theta)$])
  for t in range(100): scene = step(scene)
  return scene
\end{lstlisting}
Figure~\ref{fig:sim-demo}~shows some sample simulations generated by this function. Notice that by varying $\theta$ and $\iota$, we can produce both ``jumping'' and ``walking'' actions.

Now, let us write a memo model that reasons about the agent by interfacing with this physics simulation. In our model, the \lstinline{subject} in \citeauthor{csibra2003one}'s experiment is unsure whether or not a wall is present. However, the subject thinks that the \lstinline{agent} knows whether the wall is present. The subject models the \lstinline{agent} as a rational actor who, by the principle of rational action \citep{jara2016naive}, chooses an action $(\theta, \iota)$ that maximizes the agent's utility. We model the agent's noisily-rational choice via the softmax choice rule \citep{luce1959individual, franke2023softmax}.
\begin{lstlisting}
subject: thinks[
  world: given(wall in Bool, wpp=1),
  agent: knows(world.wall)
  agent: chooses(
    $\theta$ in Angle, $\iota$ in Impulse,
    wpp=exp($\beta$ * utility($\theta$, $\iota$, world.wall))
  )
]
\end{lstlisting}
What \emph{is} the agent's utility? Utilities are given by costs and rewards. Here, let us say that the cost of an action is proportional to the magnitude of impulse needed, $\iota$, and the reward is given by whether or not the agent is at the goal after the simulation. We can implement this utility as a Python function.
\begin{lstlisting}
def utility(wall, $\iota$, $\theta$):
  scene = simulate(wall, $\iota$, $\theta$)
  reward = 1 if agent_at_goal(scene) else 0
  cost = $\iota$
  return reward - cost
\end{lstlisting}
We can reference the function \lstinline{utility} from our memo model because memo's FFI allows direct calls to Python functions. The function \lstinline{utility} in turn calls \lstinline{simulate} through JAX2d's API. This is how memo ``talks to'' JAX2d. Notice that it is not obvious that this should be possible at all: memo was not explicitly designed to be compatible with JAX2d, and JAX2d was not explicitly designed to be compatible with memo. In fact, these software packages were developed in isolation by teams that never communicated. Nonetheless, these individually-developed domain-specific systems are compatible because of their FFIs and APIs.

Let us now complete our model. The subject observes the agent's chosen action, and then makes an inference about the presence of the wall.
\begin{lstlisting}
subject: observes [agent.$\theta$]
subject: observes [agent.$\iota$]
return subject[Pr[world.wall]]
\end{lstlisting}
Consistent with adult intuitions, this model predicts the existence of a wall if the agent jumps.

This way of modeling can be extended to reason about more complex scenarios as well. For example, we can build a memo-and-JAX2d model of \citet{liu2017ten}'s studies,\footnote{See \url{https://github.com/kach/memo/blob/main/demo/demo-liu.ipynb}.} which extend \citeauthor{gergely1995taking}'s paradigm by varying the height of the wall and which of two possible targets the agent approached. Subjects were first familiarized with four events, two per target. For one target (``low-value''), the agent jumped over a low wall, but declined to jump over a medium wall. For the other target (``high-value''), the agent jumped over a medium wall, but declined to jump over a high wall. After these four familiarization events, subjects saw a test event where the agent chose one of these two targets to approach. Subjects expected the agent chose the ``high-value'' target over the ``low-value'' target. Our model makes the same prediction, first inferring from the four familiarization events which target the agent prefers, and then predicting the agent's behavior in the test event.

% \shortcites{netanyahu2021phase, shu2021agent, liu2017ten, ullman2009help}
To be clear, there is nothing particularly novel about creating a model that makes such predictions. Many existing models can perform social reasoning grounded in physical interactions \citep[see, e.g.][]{netanyahu2021phase, shu2021agent, ullman2009help}. But these models are typically thousands of lines of code long, densely interleaving implementations of physics simulations and forward- and inverse-planning algorithms in a general-purpose programming language. What is interesting about our model is its construction out of two modular components. Our JAX2d program modeled the agent's physical body, without regard to psychology, and our memo program modeled the agent's mind, without regard to physics. Importantly, we did not have to implement \emph{theories} of physics or psychology---we never directly expressed Newton's laws or Bayes' rule---because these theories were already implemented in the respective compilers of JAX2d and memo.

%In our model, intuitive psychology makes a call to intuitive physics. But we could just as easily have connected these systems in the reverse direction, where intuitive physics makes a call to intuitive psychology. For example, we can build a model of psychology-dependent ``cartoon physics,'' where an agent runs off a cliff but continues to be supported until they look down, at which point they begin to fall. To build this model, we would have the JAX2d code decide whether to apply gravity to the body by querying memo code for the agent's mental state. We can also nest psychology and physics deeper than two levels: for example, we can build a model of an agent watching a cartoon character and making an inference about what the cartoon character knows. In this way, multilingual programming gives a blueprint for a mind that flexibly composes multiple intuitive theories.

\subsection{Assembly instructions for the modular mind}\label{sec:final-assembly}

Thus far in this section, we have offered an explanation of how a domain-specific system could in principle be supplied with exogenous world knowledge. But we have not yet addressed how the mind as a whole decides \emph{which} exogenous world knowledge to supply to a domain-specific system. We sidestepped this problem by hand-curating the world knowledge that was supplied to the model: for example, in the food trucks model, we hand-wrote the definitions of $S$, $A$, and \lstinline{take_a_step} up-front. But let us now consider where this world knowledge might come from. When faced with the food trucks problem, how does the mind determine that the model should include knowledge of the impermeability of brick walls, but not (for example) knowledge of brick chemistry or bricklaying patterns \citep[the ``frame problem'' of][]{mccarthy1981some}? And, for that matter, how does the mind decide to apply the domain-specific system for intuitive psychology in the first place?

One candidate answer to these questions comes from a recent line of work by \citet{wong2025modeling} and \citet{brooke2023bounded}, who, like us, adopt the metaphor of thinking as writing and executing programs in a mental programming language. They propose that when the mind is faced with a new problem, it first draws on a global associative store to surface the relevant world knowledge, and then applies that world knowledge to program a situation-specific model to reason over. \citeauthor{wong2025modeling} offer a concrete instantiation of this ``model synthesis architecture'' (MSA) by using a pretrained large language model (LLM)---a natural choice, because LLMs can function both as a global store of world knowledge (because they are trained on large swathes of the Internet), and as proficient program synthesis systems (because they are trained specifically on code). When faced with a particular problem, the MSA uses the LLM's language-to-code facilities to program a situation-specific model, which in turn can be executed to generate predictions and inferences about the situation at hand. (Notice that this strategy is different from directly using an LLM to generate predictions and inferences, without instantiating any intermediate programmatic representation of the problem.) The MSA generates human-like predictions about a variety of different open-world scenarios. For example, by drawing on information about sports encoded in the LLM's weights, the MSA can generate models to reason about the outcomes of arbitrary sporting events on demand, without requiring the model designers to curate bespoke knowledge of those particular sporting events ahead of time.

Can an MSA be extended to recruit, not just isolated bits of world knowledge, but an entire intuitive theory, to solve a problem? On our account, an MSA can choose to apply an intuitive theory by choosing to program its model, not in a general-purpose programming language, but rather in the theory's respective DSL. For example, when faced with the food trucks problem, the MSA may determine by association that intuitive psychology is a relevant body of knowledge to bring to bear in solving the problem. It would then choose to deploy theory-of-mind by programming a model in memo, and it would supply the model with the requisite exogenous world knowledge in the manner described earlier in this section.

As a proof of concept, we developed a simple MSA that is capable of writing memo code. The MSA is implemented as an LLM\footnote{For this demonstration, we used DeepSeek v4 \citep{xu2026deepseek}, the most sophisticated open-weight LLM available at the time of writing. The scientific benefit of using an open-weight model, rather than a proprietary model, is that we can run and inspect the entire model ourselves instead of relying on black-box queries to the servers of AI companies.} equipped with a brief system prompt, which instructs the LLM to solve the problem by deciding which programming language(s) to use and then generating the appropriate code. The MSA is also supplied with a copy of the official memo instruction manual.

To see this MSA in action, consider the following problem:
\begin{quotation}
Travis lives in San Diego. One day he decides to visit his friend Rachel, who is an undergraduate at a school in the University of California system. Travis looks up where Rachel goes to school, and then he buys a plane ticket. Does Travis think Rachel studies at UC Irvine?
\end{quotation}
\begin{figure}
\begin{lstlisting}
class School:
    UCSD = 0      # UC San Diego
    UCLA = 1      # UC Los Angeles
    UCB = 2       # UC Berkeley
    UCD = 3       # UC Davis
    UCSB = 4      # UC Santa Barbara
    UCI = 5       # UC Irvine
    UCSC = 6      # UC Santa Cruz
    UCR = 7       # UC Riverside
    UCM = 8       # UC Merced

class ModeOfTravel:
    DRIVE = 0
    TRAIN = 1
    PLANE = 2

# Table of utilities, columns = [DRIVE, TRAIN, PLANE]
U = array([
    [5, 2, -5],   # UCSD
    [4, 3, -3],   # UCLA
    [1, 0, 5],    # UCB
    [1, 0, 4],    # UCD
    [3, 1, 2],    # UCSB
    [4, 2, -2],   # UCI
    [1, 0, 3],    # UCSC
    [4, 1, -3],   # UCR
    [1, 0, 2]     # UCM
])

@memo
def infer_goal():
    observer: thinks[
        rachel: chooses(s in School, wpp=1),
        travis: knows(rachel.s),
        travis: chooses(a in ModeOfTravel, wpp=exp($\beta$ * U(rachel.s, a)))
    ]
    observer: observes_that [travis.a == {ModeOfTravel.PLANE}]
    return observer[Pr[rachel.s == {School.UCI}]]
\end{lstlisting}
    \caption{When faced with a problem that requires mental state reasoning, our MSA marshals the relevant world knowledge from an LLM's weights, and then generates memo code that performs theory-of-mind reasoning using that world knowledge. This code was generated directly by our MSA when given the problem described in Section~\ref{sec:final-assembly}.}
    \label{fig:msa-output}
\end{figure}
When given this problem text (and no other information), the MSA automatically constructs the situation-specific program shown in Figure~\ref{fig:msa-output}, which predicts that if Travis is flying, he is likelier to be headed to UC Berkeley or UC Davis than to UC San Diego or UC Irvine, and thus concludes that he probably does not think Rachel studies at UC Irvine. Drawing on its world knowledge, the MSA first assembles the list of schools in the UC system, the possible alternative modes of transportation that Travis could have taken, and the cost of each mode for each school (which itself involves knowledge about geography, transit options in California, etc.). It then writes a memo program that applies this world knowledge to infer Travis' goal, assuming that he chose his action rationally. Notice in particular that the critical Bayesian computation, inferring Travis' goal given his action, is \emph{not} performed by feedforward passes through the LLM---it is performed by evaluating the memo program. The memo compiler itself has no built-in knowledge of the UC system or plane tickets; it is truly a domain-specialized system.

To be clear, the MSA is not specialized for reasoning about the UC system and plane tickets. The same MSA could be applied to quite different problems that require different bits of world knowledge. For example, we could just as well have posed the following problem:
\begin{quotation}
Travis lives in Boston. One day he decides to visit his friend Rachel, who is an undergraduate at an Ivy League university. Travis looks up where Rachel goes to school, and then he buys a train ticket. Does Travis think Rachel studies at Dartmouth?
\end{quotation}
To solve this problem, the MSA will generate code that resembles Figure~\ref{fig:msa-output}, but with different world knowledge: that is, different possible goals, actions, and utilities. On the other hand, if we pose a problem that does not require reasoning about the mental states of agents, the MSA does not write any memo code at all. For example, consider the problem:
\begin{quotation}
Is it faster to drive from Boston to Dartmouth, or Boston to Yale?
\end{quotation}
Here, the MSA draws on its world knowledge to estimate driving times to Dartmouth and Yale, but it then uses that knowledge to generate ordinary Python code, not memo code, because there is no need to reason about other agents' mental states (Figure~\ref{fig:msa-no-memo}).

\begin{figure*}
\begin{lstlisting}
# "Is it faster to drive from Boston to Dartmouth, or Boston to Yale?"
def compare_travel_times():
    dartmouth_driving_time = 2.0   # hours
    yale_driving_time = 2.5        # hours
    return dartmouth_time < yale_time
\end{lstlisting}
\caption{If mental state reasoning is not required to reason about a situation, then our MSA does not generate any memo code. Instead, it reasons using ordinary general-purpose Python code. (Compare to Figure~\ref{fig:msa-output}.)}\label{fig:msa-no-memo}
\end{figure*}

In this way, our MSA shows, in principle, that it is possible for a cognitive system to solve open-world problems by drawing on its world knowledge, and, when necessary, calling on a modular, domain-specialized system. Our prototype serves as a proof of concept that it is possible to combine dynamically-surfaced domain-general knowledge with a domain-specialized system. (For a more sophisticated demonstration in this spirit, see \citet{ying2025language}.)

We can think of our MSA as a computational instantiation of \citet{casto2025does}'s ``exportation of information hypothesis'': that when faced with linguistic input, the brain's core language system extracts and exports information to functionally-specialized brain regions for further processing. In fact, evidence for this hypothesis is strongest in the domain of theory-of-mind: in fMRI experiments, the theory-of-mind network responds strongly when participants read stories that require reasoning about mental representations, but not when participants read stories that require reasoning about physical representations  \citep{saxe2013people}; in contrast, the language network responds equally to both types of stories \citep{shain2023no}. This is consistent with the hypothesis that the language network extracts information that is selectively exported to the theory-of-mind network when applicable, architecturally analogous to how the LLM extracts and exports information to memo in our prototype MSA.

\subsection{Why have domains at all?}

Let us summarize our answer to \citeauthor{fodor1983modularity}'s puzzle. How is it possible to have a system for social cognition that, on one hand, is domain-specialized, and, on the other hand, can still solve real-world problems? On our account, DSLs offer a way to formalize a theory of a domain in a way that is informationally encapsulated, but nonetheless allows the programmer to supply incoming exogenous world knowledge. The FFIs and APIs of separate DSLs allow them to interface with each other, explaining how the mind might combine different intuitive theories. Finally, MSAs are a candidate model of how, when faced with a problem, the mind might surface and apply the relevant world knowledge and intuitive theories in the first place.

The demonstrations we presented in this section raise one final question about domain-specialized systems. If an MSA can surface any relevant knowledge to construct a situation-specific model, then why choose to use a DSL at all? What is the value of carving up knowledge into domains?

Reasoning by analogy to the value of real-world DSLs, we propose that the division of cognition into domains is ultimately valuable for cognitive efficiency.
The advantage of using a DSL as the target language is that knowledge of the domain is included for free in the DSL's syntax and semantics, and need not be surfaced and implemented afresh in every new program.
Simply by choosing to use memo, the MSA automatically activates knowledge of the principle of rational action and the referential opacity of belief. It is freed from the responsibility of programming these computations from scratch and ensuring their correctness every time it is faced with a new problem. This suggests two virtues of using a domain-specialized system: first, it takes less work to generate a given model, and second, the generated models are more likely to be correct (consistent with the body of knowledge that comprises the theory).

To observe a third virtue of domain-specificity, notice that DSLs also hold the dual property to encapsulation, what \citeauthor{fodor1983modularity} calls ``inaccessibility.'' If encapsulation means that the system can only draw on limited external information, inaccessibility means that external processes can only access limited information from within the system. For example, a SQL programmer cannot access the internal planning mechanism the SQL compiler uses to efficiently execute database operations, and a memo programmer cannot access the internal data structures that the memo compiler uses to efficiently run model inference. This limited access serves an important purpose: it allows designers of DSLs to freely change the internal representations used by the compiler, without worrying about breaking programs that depend on those internal representations \citep[\S5]{ousterhout2018philosophy}. For example, SQL and memo both routinely receive updates that make programs run faster. These updates sometimes radically restructure the compiler's internal data structures, but, because the syntax and semantics of the language remain the same, users (and MSAs) do not even notice that the compiler's internals have changed. This feature of DSLs suggests a third virtue of domain-specialized cognitive systems: they allow the rest of cognition to immediately and automatically benefit from improvements to that system, e.g.\ from a more efficient algorithmic-level implementation of a given computation.

In these ways, carving knowledge into domains simplifies the computational task of generating efficient and correct situation-specific models to reason about new problems.
\section{The space of possible theories of mind}\label{sec:memo-junior}

\begin{figure*}
\centering
\includegraphics[width=\linewidth]{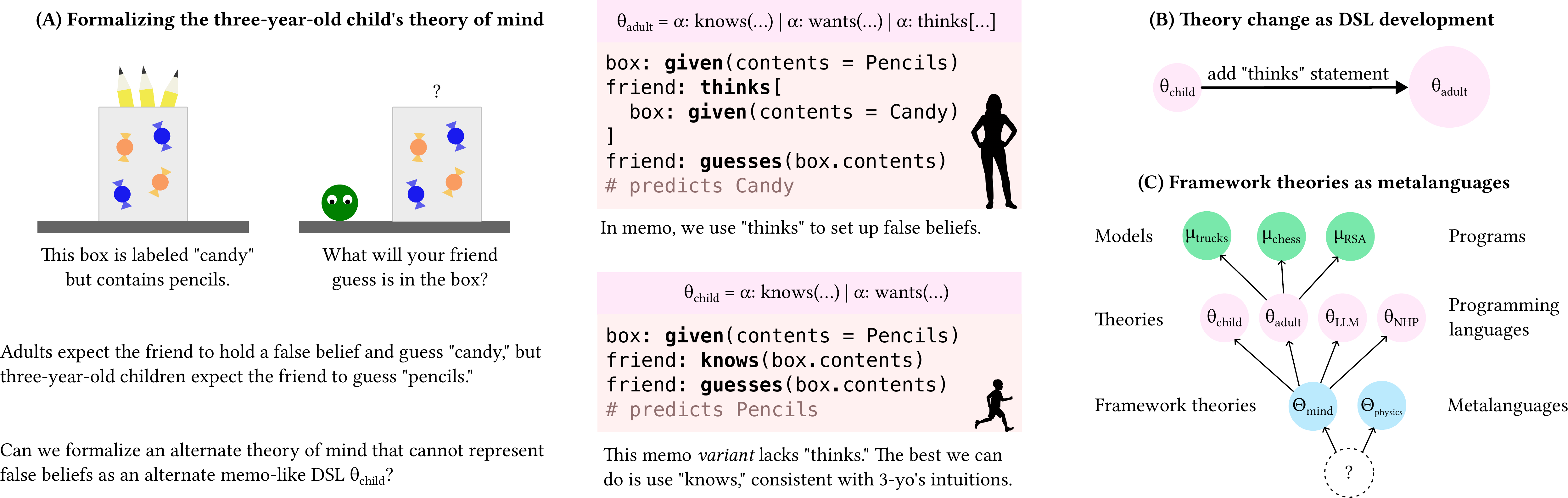}
\caption{How can we formalize alternate theories of mind? (A) In Section~\ref{sec:memo-junior}, we take as a case study the theory of mind of a three-year-old child, and show how it can be formalized as a variant of memo that lacks the construct \lstinline{thinks}. (B) Next, we discuss how we can cast theory change in terms of the development of a DSL (\S\ref{sec:theory-change}). (C) Finally, we discuss how we can think of framework theories in terms of meta-DSLs that are themselves designed for implementing DSLs (\S\ref{sec:framework-theories}).}\label{ref:3-summary-fig}
\end{figure*}

We have argued so far that $\theta$, the theory-of-mind of a typical human adult, can be modeled by the DSL memo. But our goal in this paper is not only to model one particular theory-of-mind $\theta$, but also to articulate alternate possible theories of mind: the third challenge raised in the introduction.

To make progress on this goal, let us consider a case study: modeling a particular stage of theory-of-mind development. There are many stages and developmental milestones we could study \citep{wellman2011sequential}, but here we will focus on one of the most well-known examples: the development of false belief understanding. While much remains disputed in the theory-of-mind development literature, it is at least broadly agreed upon that three-year-old children systematically fail tasks that require them to represent the false beliefs held by other agents \citep{baron1985does, wimmer1983beliefs}. Let us stipulate for the moment that this failure really is due to three-year-old children being unable to represent false beliefs. Call the three-year-old child's theory-of-mind $\theta_3$. On our account, it should be possible to create a variant of memo that formalizes $\theta_3$ rather than $\theta$. But what might this ``variant'' of memo look like?

The notion of ``language variants'' is well-studied in programming language theory. For example, some computer science curricula introduce programming through a progression of increasingly-powerful ``student languages'' that progressively introduce more syntactic forms \shortcites{felleisen2018design}\citep{tsur2018scratch, felleisen2018design}. Language variants can also be used as executable models of students' \emph{misconceptions} about how a language works \shortcites{chandra2024watchat}\citep{chandra2024watchat, lu2024identifying}.

\begin{figure}
\begin{bnf}  % https://ctan.math.washington.edu/tex-archive/macros/latex/contrib/simplebnf/simplebnf-doc.pdf
$\langle$\emph{statement}$\rangle$ ::=
| $a\colon$ chooses($x$ in $D$, wpp=$\langle$\emph{expression}$\rangle$)
| $a\colon$ knows($b.x$)
| $a\colon$ observes $b.x$ is $c.y$
| {\color{red}$a\colon$ thinks [$\langle$\emph{statement}$\rangle\dots$]}
;;
\end{bnf}
%$\text{alice}\llbracket \; \text{observes} \; x \;\;\; \mathclap{\leftarrow}\rrbracket\;\;y$
\caption{Syntax of statements in memo (abridged). Variables $a, b, c$ range over names of agents, and $x, y$ over names of choices. Notice that the \lstinline{thinks} statement has a uniquely recursive syntactic structure that itself references the $\langle$\emph{statement}$\rangle$ rule. By omitting this rule from the syntax, we obtain a language that cannot represent false or accidentally-true beliefs.}\label{fig:grammar}
\end{figure}

To construct a variant of memo that formalizes $\theta_3$, let us start by recalling from Section~\ref{sec:structural-variation} that in memo, an agent's beliefs are represented using the \lstinline{thinks} statement. We have already seen how to use \lstinline{thinks} to represent agents' true beliefs and uncertainty. For example, we modeled Zoe seeking Ali based on her mental model of Ali's choice of bar. But in general, the \lstinline{thinks} construct can also be used to represent an agent's \emph{false} belief about the world.

Suppose that Zoe had a false belief about Ali's location. Perhaps Zoe erroneously thinks that Ali chooses a cheap bar, when in reality he actually chooses a nearby one. To model this situation in memo, we might write:
\begin{lstlisting}
Ali: chooses(b in Bar, wpp=exp(quality(b) - distance(b)))
Zoe: thinks[
  Ali: chooses(b in Bar, wpp=exp(quality(b) - price(b)))
]	
\end{lstlisting}
Here, Zoe's model of Ali's choice differs from his actual choice. In this way, \lstinline{thinks} allows representing false beliefs.

In practice, it turns out that the full generality of the \lstinline{thinks} statement is typically unnecessary. Most of the time, memo programmers simply need to express that an agent correctly knows another agent's choice. Suppose Zoe knows ahead of time exactly where Ali is, perhaps because he had posted an update on social media. How could we build this model? One approach is to write that Zoe \lstinline{thinks} Ali chooses a bar and then correctly \lstinline{observes} where he really went:
\begin{lstlisting}
Ali: chooses(b in Bars, wpp=exp(quality(b) - distance(b)))
Zoe: thinks[
  Ali: chooses(b in Bars, wpp=exp(quality(b) - distance(b)))
]
Zoe: observes Ali.b is Ali.b
\end{lstlisting}
This model sets up the correct mental state, but it is somewhat dissatisfying in its implementation. It is verbose and redundant, requiring the programmer to re-express Ali's choice in Zoe's mental model. Because such situations are so common in everyday modeling, memo provides a shorthand for these situations: the \lstinline{knows} statement, which asserts that an agent knows the exact value of some variable they otherwise would not have access to. Using \lstinline{knows}, we can express this situation more succinctly:
\begin{lstlisting}
Ali: chooses(b in Bars, wpp=exp(quality(b) - distance(b)))
Zoe: knows(Ali.b)
\end{lstlisting}

Suppose that we created a variant of memo that dispenses with the complex \lstinline{thinks} and \lstinline{observes} statements, but retains everything else, including the simpler \lstinline{knows} statement. What would we lose? The resulting programming language, which we might call ``memo-junior,'' could still be used to model a wide variety of scenarios. However, memo-junior would not be able to represent scenarios where agents have false beliefs. If memo variants formally encode theories of mind, then memo-junior might be closer to encoding the three-year-old child's theory of mind than the adult's.

Let us work through the idea that memo-junior formalizes $\theta_3$, the three-year-old child's theory of mind. As an example, consider how we might model the classic false-contents ``Smarties'' task in memo and memo-junior \citep{perner1987three}. In this task, a child is given a candy box and shown that it contains a pencil, not candy. The child is then asked to predict whether a friend waiting outside the room will expect the box to contain a pencil or candy. At age 3, children systematically mis-predict that the friend will expect the box to contain a pencil.

In the full version of memo, it is straightforward to represent that the friend has a false belief that the box's contents match its label, i.e.\ that the box contains candy. The friend then makes a guess to maximize the chances of guessing correctly.
\begin{lstlisting}
# observer's belief determined by actual contents of box
box: given(contents, such_that=(contents == {Pencil}))
friend: thinks[
  # friend's belief determined by label on box
  box: given(contents$^\prime$, such_that=(contents$^\prime$ == {Candy}))
]
friend: chooses(guess, to_maximize=Pr[guess == box.contents$^\prime$])
\end{lstlisting}
In this model, the (unprimed) variable \lstinline{box.contents} refers to the observer's belief about the box's actual contents, while the (primed) variable \lstinline{box.contents$^\prime$} refers to the observer's meta-representation of the friend's belief about the box's contents. Consistent with adult observers' intuitions, this model predicts that the friend will incorrectly guess that the box contains candy, not a pencil. In memo-junior, however, we cannot use the \lstinline{thinks} statement to write this model. If we are forced to generate a prediction for the friend's guess, the closest we can get to writing a model that compiles is by using \lstinline{knows} to give the friend direct access to the observer's own representation of the box's contents.
\begin{lstlisting}
box: given(contents, such_that=(contents == {Pencil}))
friend: knows(box.contents)  # refers directly to observer's variable
friend: chooses(guess, to_maximize=Pr[guess == box.contents])
\end{lstlisting}
Consistent with the intuitions of three-year-old children, this memo-junior model predicts that the friend will correctly guess that the box contains a pencil.

\subsection{What empirical evidence is memo-junior consistent with?}\label{sec:memojrempirical}

memo-junior restricts the space of possible models that can be written, in a way that is consistent with empirical evidence about the intuitions of three-year-old children on false belief tasks. In fact, the design of memo and memo-junior is consistent with several other pieces of empirical evidence, too.

First, the \lstinline{knows} statement in memo and memo-junior is along many dimensions more fundamental than the \lstinline{thinks} and \lstinline{observes} statements that are added in the full version of memo. The syntax of \lstinline{knows} is simpler than the compound, recursive structure of \lstinline{thinks} (see Figure~\ref{fig:grammar}). In real-world memo code \lstinline{knows} is used more frequently than either \lstinline{thinks} or \lstinline{observes}: among the 24 memo models published in memo's documentation at the time of writing, \lstinline{knows} was used 87 times while \lstinline{thinks} and \lstinline{observes} were used 43 and 39 times, respectively. The \lstinline{knows} statement requires fewer lines of code to implement in the memo compiler than \lstinline{thinks} and \lstinline{observes}. Finally, \lstinline{knows} runs faster and with less memory usage than equivalent code written using \lstinline{thinks} and \lstinline{observes}, because it need only construct a lightweight pointer relation to the known information rather than a fresh representation of that information. These properties are consistent with the many converging lines of evidence \citet{phillips2021knowledge} marshal to argue that knowledge conceptually precedes belief in humans: that rather than thinking of knowledge as a special case of belief (``justified true belief''), we should take knowledge to be the more fundamental building-block that precedes the more advanced notion of belief.

Second, memo and memo-junior are consistent with the observation that, even before the development of false belief understanding, children seem to have a cogent intuitive psychology that integrates concepts of desire, emotion, and perception \citep[see][Ch.\ 4]{wellman2014making}. For example, before children can represent others' beliefs, they can represent that others have distinct desires, and that those desires influence action and emotion \citep{repacholi1997early, wellman1990simple, wellman2004scaling}. Consider \citet{repacholi1997early}'s classic broccoli--goldfish task: when an adult expressed disgust at Goldfish crackers and appeal at raw broccoli, children who themselves preferred the crackers chose to give the adults broccoli, indicating that they understood that the adults' preferences differed from their own, and that they could infer the adults' preferences from their emotional displays. We can model children's behavior in this task using memo-junior: the model does not require the child to use \lstinline{thinks}.\footnote{See \url{https://github.com/kach/memo/blob/main/demo/demo-broccoli.ipynb}.} In this way, memo-junior is consistent with the behavior of children beyond just their failure on false belief tasks.

A number of philosophers have noted a puzzle in the fact that desire attribution emerges much earlier than belief attribution \citep{rakoczy2007way, steglich2015desire, dealwis2025desire}. The puzzle is that beliefs and desires are seemingly symmetric: representing either beliefs or desires requires representing a potentially-non-actual state of the world, either the believed present state or the desired future state \citep{anscombe1957intention}. Why, then, does desire attribution emerge before belief attribution? \citet{foster2025desire} argue that theory-of-mind development exhibits this asymmetry because representing belief requires meta-representation, while representing desire does not. As we have suggested elsewhere \citep{chandra2026conniving}, memo and memo-junior are consistent with this account of the data. The distinct recursive structure characteristic of \lstinline{thinks} is required for representing beliefs, but not desires. How is this possible? In both memo and memo-junior, desires are represented using a statement called \lstinline{wants}. Unlike \lstinline{thinks}, \lstinline{wants} does not require programmers to explicitly recursively embed a possible world state. Instead, \lstinline{wants} automatically derives possible future world states by looking ahead in the program to see what might happen next---an operation that is formalized by the well-studied theory of \emph{continuations} in programming languages \citep{appel2007compiling, haynes1984continuations, reynolds1993discoveries}. Hence, a three-year-old child can represent desires using \lstinline{wants} before she invents the recursive syntax needed to represent beliefs with \lstinline{thinks}.

\subsection{Beyond $\theta$ and $\theta_3$}\label{sec:beyondtheta}

In this section, we have taken $\theta$ and $\theta_3$ as case studies of theories of mind that can be formalized as DSL variants. Before we continue, let us note that this methodological approach allows us to begin formalizing many other theories-of-mind beyond $\theta$ and $\theta_3$. For example, we could start designing memo variants that formalize our hypotheses of the theories-of-mind held by aging minds, non-human primates \citep[see][for recent work on formalizing NHP theories-of-mind]{horschler2023differences}, and potentially even AI systems \citep[see][]{hu2025re}. These formalizations in turn would precisely specify how we think these theories-of-mind differ, committing to predictions that could be tested empirically.

Furthermore, this approach allows us to express not only the theories-of-mind held by minds, but the theories-of-mind that minds attribute to \emph{other} minds: for example, how an adult might think about a child, a machine \citep{brink2020robot}, or an omniscient being \citep{barrett2001god, lane2010children}. These formal models can also be used as hypotheses in higher-level models of how minds decide \emph{which} theory-of-mind to attribute to an agent they observe in the world \citep{gray2012mind}; see \citet{burger2020mental} for early models in this spirit.

\subsection{Theory change as DSL development}\label{sec:theory-change}

Having formalized both the pre-theory ($\theta_3$) and the post-theory ($\theta$), we can begin to study why and how a child might transition from the former to the latter.\footnote{We will continue using the development of false belief understanding as a case study in conceptual development, but the same approach could in principle be applied to study other instances of conceptual change in intuitive psychology: for example, how children come to understand sophisticated emotion concepts \citep{nook2019emotion}, or even how adult clinical psychologists develop their understanding of the mind through training and practice \citep{luhrmann2011two}.} As \citet[pg.~20]{carey2009origin} writes, to give a full account of conceptual development, we must provide two things: first, ``a satisfactory characterization of what it means for a representational system to be qualitatively different from, to transcend, those that preceded it,'' and second, ``a learning mechanism that accounts for how new representational capacity could come into being.'' The DSL account can help us make progress on both of these challenges.

First, treating intuitive theories as DSLs allows us to precisely articulate claims about the expressivity and incommensurability of theories, in terms of whether DSL variants are mutually compatible. For example, the full version of memo is strictly an extension of memo-junior: hence, all memo-junior models are valid memo models, but not vice versa, and there exist memo models that cannot be translated into memo-junior models (such as our model of the false-contents task above). This strict superset relationship between memo and memo-junior is one sense in which we may formally characterize memo as ``being qualitatively different from, and transcending'' memo-junior.

Second, treating conceptual development as DSL development can help us understand how intuitive theories gain expressive power. Most immediately, DSLs offer a representational format over which a learning mechanism could operate. For example, recall that \citet{gopnik1997words} liken the dynamics of theory change to scientific theory change \citep{kuhn1962structure, lakatos1970history}. Building on \citet{gopnik2004theory}'s Bayesian-network account, \citet{goodman2006intuitive} provided a computational model of theory-of-mind development as hierarchical inference over a space of possible Bayesian networks. In this model, as evidence accrues, a learner transitions to a more sophisticated belief-desire theory because the predictive power of a more sophisticated ``belief-desire'' network eventually outweighs its \emph{a priori} complexity. As we have previously discussed, Bayesian networks cannot give a full account of theories of mind. However, the same modeling approach could be instantiated with DSLs instead of Bayesian networks. On our account, we might model theory change as hierarchical Bayesian inference over a hypothesis space of \emph{candidate DSLs} of varying complexity and expressivity. For example, a child who is considering the candidate construct \lstinline{thinks} might notice that models written using \lstinline{thinks} in memo predict people's behavior better than models written without \lstinline{thinks} in memo-junior, and thus decide rationally to invest in using this more complex construct. (The next section speculates on what the full hypothesis space of candidate DSLs might be.)

How might proposals for new candidate constructs like \lstinline{thinks} originate in the first place? One source of new concepts in a theory of mind may be natural language.
Many researchers have argued that learning natural language supports theory-of-mind development \citep{astington2005language, milligan2007language, bartsch1995children, astington1999longitudinal, de2000linguistic, pyers2009language, peterson2000insights, harris2005language}. If we treat theory change as DSL development, then we can model how a child might use natural language input to learn new mentalistic concepts in terms of discovering and bootstrapping new syntactic constructs, such as \lstinline{thinks} in the transition from memo-junior to memo. For an analogous modeling effort in the domain of numerical concept learning, see \citet{piantadosi2012bootstrapping}, and, more recently, \citet{das2026one}.

Another possible source of new concepts may be an endogenous search for useful abstractions.
\citet{ellis2021dreamcoder} describe a formal model that grows DSLs by recognizing common programming patterns across a suite of existing programs, and then formalizing those patterns into reusable new constructs within the DSL. The model can then use these constructs to write new programs that would previously be too complex to write. As \citet{rule2025end} argue, these new constructs increase the DSL's expressive power, at least for a resource-bounded agent. If we think of theories-of-mind as DSLs, we can imagine applying such a system to formally model how we might grow our intuitive theories pragmatically, in the sense of \citet{peirce1878how}: that is, how the process of theory development may be driven by the need to expand representational efficiency and predictive power for the tasks we encounter in the world. This is, after all, how real-world programming languages tend to grow, too: not from purely theoretical innovations, but from demonstrated utility for solving ecological problems \citep{meyerovich2012socio, meijer2007confessions, stroustrup2020thriving}.
A formal model of this type of process could help us understand why our theories of mind seem to be architected around abstractions like ``beliefs'' and ``desires'' rather than any other types of abstractions.

Finally, treating intuitive theories as DSLs might help make sense of the puzzling dynamics of theory change. Theory change is a messy business. \citet{gopnik1997words} describe the liminal period when a new theory is gingerly considered, but the old theory is not yet abandoned. Indeed, \citet{amsterlaw2006theories} find that a child on a given day might show evidence of using two or three different theories-of-mind across tasks. Interestingly, PL design follows the same pattern: growing programming languages is a messy business, too \citep{steele1998growing, gabriel2012structure}. As an example, consider the case of the Python programming language. In the early 2000s, it became increasingly clear to Python users that the abstractions for working with text (``strings'') were fundamentally flawed---the language failed to make the critical distinction between a sequence of characters and a sequence of bytes representing those characters. This made it difficult to process text input written in non-Latin alphabets, a serious problem for an increasingly global population of programmers. The authors of the language thus redesigned these abstractions from scratch in the next major version, Python~3, which was released in 2008 \citep{rossum2008whats}, and was not compatible with Python~2. We can think of this new version of the language as a kind of ``paradigm shift'' (in \citeauthor{gopnik1997words}'s terms), or as an ``incommensurable change'' (in \citeauthor{carey2009origin}'s terms) in how text was handled.

Indeed, just like with scientific paradigm shifts, the transition from Python~2 to Python~3 was not immediate \citep{coghlan2012python}. The old version, Python~2, remained actively used for the next twelve years, and was only formally phased-out in 2020. There were many reasons programmers hesitated to switch to Python~3 in the interim: some did not want to rewrite their existing Python~2 code in Python~3, while others depended on libraries only available in Python~2. Hence, for many years the choice between Python~2 and Python~3 was a pragmatic one: programmers embarking on new projects, or who really needed Python~3's features, generally chose the new version, while programmers working with large existing codebases continued to use Python~2 as long as they could.
%Many programmers even wrote their code carefully to be compatible with both versions at once \citep{malloy2019empirical}.
We might expect a similar dynamic to govern the adoption of new theories-of-mind by children. Even though a new theory might be compelling, there might be pragmatic value in keeping around an older, tried-and-tested theory---perhaps to recycle models and computations that are only compatible with the old theory. A child might thus show patterns of behavior that, to an observing scientist, seem inconsistent with any one theory-of-mind.

\subsection{Framework theories and $\Theta$}\label{sec:framework-theories}

The theories-of-mind $\theta$ and $\theta_3$ are both part of some large set $\Theta$ of possible theories-of-mind. While we have offered a way to formalize individual theories $\theta_i \in \Theta$ in terms of variants of memo, we have not yet offered a language in which we could formally describe $\Theta$ itself.

This observation suggests adding a fourth challenge to the list presented in the introduction to this paper. This fourth challenge is analogous to the first three challenges, but up one level of abstraction: the challenge is to present a mathematical framework for formalizing theories of mind that allows us to also formalize the space $\Theta$ of possible theories-of-mind, of which $\theta$ and $\theta_3$ are two members. In the same way that the object $\theta$ represents what is shared across all models $\{\mu_1, \mu_2, \mu_3, \dots\}$, the object $\Theta$ represents what is shared across all theories-of-mind $\{\theta, \theta_3, \dots\}$. In this way, $\Theta$ would formalize the domain of intuitive psychology, allowing us to determine (for example) whether a given theory is a theory-of-\emph{mind} or not by testing whether or not that theory is a member of $\Theta$. Finally, our mathematical framework should allow us to specify what kind of thing $\Theta$ itself is; that is, what the space of possible domains is.

We have not yet developed a concrete answer to this fourth challenge. But our approach to the first three challenges, treating $\theta$ and $\theta_3$ as programming languages, suggests a possible approach to the fourth challenge: formalizing a space of possible domain-specialized programming languages.

There are many ways to approach the task of formalizing the ``metatheory'' of a space of programming languages. For example, we could begin specifying the space of possible grammars for a DSL by creating a meta-grammar that generates grammars. However, in keeping with the DSL perspective, let us offer a different view. Notice that a DSL is typically \emph{itself} instantiated as a program: the compiler or interpreter that runs code written in the DSL is itself a piece of code. In fact, DSLs are often designed using meta-DSLs specialized for implementing DSLs. For example, it is common to express the syntax of a DSL using a meta-DSL for expressing grammars. One such meta-DSL is called ``YACC,'' which stands for ``Yet Another Compiler Compiler,'' a phrase that expresses both the abundance and recursive nature of such DSLs \citep{levine1992lex}. Other meta-DSLs are specialized for implementing particular kinds of DSLs: for example, the Rosette meta-DSL \citep{torlak2013growing} is specialized for creating DSLs that natively support program synthesis. Rosette itself is implemented using Racket, a (meta-)meta-DSL specialized for what its designers call ``language-oriented programming'' \citep{felleisen2018programmable, butterick2016beautiful}. In a sense, programming languages are ``DSLs all the way down'' \citep{thompson1984reflections}.

We can think of the relationship between meta-DSLs and DSLs as analogous to the relationship between ``framework theories'' and ``specific theories'' \citep{wellman1992cognitive}. In the same way that a framework theory constrains the space of possible specific theories, providing a strong inductive bias that guides theory change over time, a meta-DSL constrains the space of possible DSLs that can be implemented, providing a strong inductive bias that guides DSL development over time. Similar to how \citet{tenenbaum2007intuitive} imagine a hierarchy of increasingly abstract theories, from specific theories to framework theories and onward, we imagine a hierarchy of increasingly abstract DSLs that generate DSLs, meta-DSLs, and so on. On this account, then, $\Theta$ should be formalized by a meta-DSL in which one can implement any memo-like DSL.

Let us summarize: In this section, we proposed that we can formalize possible alternate theories of mind as variants of memo. As a case study, we considered the theory of mind $\theta_3$ of a three-year-old child who systematically fails tests of false belief understanding, and we suggested that $\theta_3$ could be formalized as a variant of memo where the \lstinline{thinks} statement is removed. We discussed the empirical evidence consistent with this idea. Finally, we proposed that the set $\Theta$ of possible theories-of-mind could be formalized as a meta-DSL for implementing memo-like DSL variants.

\section{Conclusions and limitations}

What is a theory of mind?
We began this paper by observing that the influential ``theory theory'' has long been contentious in part because it is not clear what exactly a theory of mind \emph{is}. What type of knowledge could be applied to reason about situations with structural variation, could be domain-specialized but able to interact with the rest of knowledge, and could be subject to conceptual revolutions over the course of development?

To address these problems, we proposed that it is fruitful to think of a theory of mind---and, indeed, any intuitive theory---as a type of domain-specialized programming language (DSL). This proposal allows us to make progress on at least three problems. First, treating a theory of mind as a DSL allows us to formalize in a fixed mathematical object the shared knowledge that supports reasoning about a wide variety of scenarios that vary structurally in the number of agents and recursive levels of reasoning. Second, a DSL allows us to formalize a theory of mind in a way that is informationally encapsulated, but can interface with external world knowledge and other intuitive theories. Third, varying the syntax and semantics of a DSL allows us to formalize and compare alternate theories of mind, and to study their development in humans as analogous to the development of real-world DSLs.

Throughout this paper, we demonstrated how this proposal could be implemented using the real-world DSL memo as a concrete worked example. Nonetheless, there are at least two ways in which our current proposal still remains a sketch, and not a complete account of theory of mind.

First, although we have shown many ways in which memo is an appealing model of theory-of-mind, we do not mean to say that memo perfectly captures a typical adult's theory of mind $\theta$. Memo is one representative member of a class of hypotheses about what $\theta$ might be, but it is not the best, nor was it explicitly designed to be. Other DSLs---including, perhaps, future versions of memo---may more precisely formalize $\theta$ than the version of memo described in this paper. Further research is required to develop alternatives, and measures for deciding which among those alternatives best formalizes theory of mind.

Second, although the current version of memo captures some core features that are widely agreed to be part of theory of mind, such as reasoning about beliefs and desires, the full scope and boundaries of theory of mind remain to be determined. For example, is the knowledge we apply to reason about social relationships and groups, cognitive resources like memory and attention, bodily states like hunger and pain, personality traits, social networks, and physical appearances part of the conceptual repertoire of theory of mind? Or is some of this knowledge part of intuitive theories of other domains, such as intuitive biology or intuitive sociology? Does some of this knowledge lie in systems that are not theory-like at all, as may be the case for the perception of faces, animacy, and social interaction?

Regardless of how we answer these questions, thinking about social reasoning in terms of DSLs offers a roadmap for formalizing and implementing the possible accounts. If we believe a concept to be part of theory of mind, we can expand memo with new features for reasoning about that concept. On the other hand, if we believe a concept to be part of an intuitive theory of a different domain, then we can use memo to interact with a separate DSL representing that domain, as we described in Section~\ref{sec:domain-specificity}.
In these ways, we hope that thinking in terms of DSLs can be clarifying and generative for researchers with a wide array of views on the nature and scope of theory of mind.

\section*{Acknowledgments}
Portions of this work were presented in non-archival form at CogSci 2025, CogSci 2026, PLATEAU 2026, and SPP 2026 \citep{chandra2025theories, chandra2026assembly}. We thank the attendees of these conferences for their feedback on this work.

\paragraph{Funding statement} This work was supported by the MIT Siegel Family Quest for Intelligence, Schmidt Sciences, the Simons Foundation, the Hertz Foundation, and a Zhou Family Fellowship.
\paragraph{Conflicts of interests statement} The authors have no conflicts of interest to declare.

\bibliographystyle{apacite}
\bibliography{refs.bib}

\newpage
\nolinenumbers
\singlespacing

\end{document}